\documentclass{article}
\usepackage{amssymb, amsmath}
\numberwithin{equation}{section}
\newtheorem{theorem}{Theorem}[section]
\newtheorem{definition}{Definition}[section]
\newtheorem{remark}{Remark}[section]
\newtheorem{proposition}{Proposition}[section]

\newtheorem{lemma}{Lemma}[section]

\begin{document}
 \title{Future $T^2$ symmetry 
  Einstein-Vlasov-Scalar field system }
\author{A. T. Lassiye$^{1}$ and D. Tegankong$^{2}$ \\
  {e-mail $^{1}$ : ltchuani@gmail.com}; \ \ \ \  
  {e-mail$^{2}$ : dtegankong@yahoo.fr}\\
  Department of Mathematics,
   Advanced Teacher Training College,\\ University of Yaounde 1, PO Box 47, Yaounde Cameroon\\ }
\date{}
\maketitle\section*{Abstract}
We prove local existence of solutions in the case of cosmological models for
 the Einstein-Vlasov-scalar field
system with $T^2$ symmetry in expanding direction. This proof is based on short-time existence theorems
for the partial differential equations resulting from the Einstein-Vlasov-Scalar Field system in areal coordinates.
 The sources of equations are generated by a distribution function and a scalar field, subject to  Vlasov
and a nonlinear wave equations respectively.\\

\textbf{Keywords} Einstein,  Vlasov,  Scalar field,  $T^2$ symmetry,
Hyperbolic differential equations,  Expanding direction, Local existence
\\
\textbf{MSC2020:} $83C05$, $83C20$, $35A01$, $35A02$, $35L40$, $35L45$, $35Q83$.
\section{Introduction}
The study of global existence Cauchy problem constitutes one of the main areas of rechearch in mathematical relativity.
An essential tool in an investigation
of this type is first of all a local in time existence theorem. A theorem of this kind is
proved here for one particular choice of matter model such as collision-less matter described by the Vlasov
equation and a nonlinear scalar field for the cosmological case.\\
In [T1] local in time cosmological
solutions of the Einstein Vlasov system with massless scalar field in surface symmetry
written in areal coordinates is obtained. The method used previously is adapted here in the case of $T^2$ symmetry.
There are several reasons why it is of interest to look at the
case of a scalar field (cf  [T1] and references therein). Spacetimes with $T^2$ symmetry have
received much attention by different authors for last years. For example in [A], [BCIM], [S] and [W], global existence result was proved in the case of Einstein-Vlasov system using  fundamental local existence in time result
(without any symmetry assumption) of Choquet [C].
But in the presence of any other matter than Vlasov, Choquet's result can not be considered. And therefore
the prove of local existence is necessary before global one.\\
The matter constituting self-gravitating physical systems
is described using certain matter fields which are also geometrical objects.
The spacetime metric is required to satisfy the Einstein equations and these
are coupled to equations of motion for the matter fields. Here, for long, the practice has been to study existence
of solutions under symmetry assumptions.
There are three types of time coordinates which have been
studied in the inhomogeneous Einstein-Vlasov system case :
constant mean curvature, conformal and areal. A constant mean
curvature  time coordinate $t$ is one where each hypersurface of
constant time has constant mean curvature and on each hypersurface
of this kind the value of $t$ is the mean curvature of that slice. In conformal coordinates the metric is
conformally flat on a two-dimensional Lorentzian manifold
which is the quotient of spacetime by the symmetry group. The time coordinate $R=t$  is the area of the symmetry orbits.
In  areal coordinates the time coordinate is a function of (or is taken to be proportional to)
the area of the surfaces of symmetry.
The paper proceeds as follows. In section 2, we present in $T^2$
symmetry expanding direction, the system in areal coordinates.
Section 3 is devoted to a priori estimations of unknowns functions and their derivatives.
Section 4 deals on local in time existence of solution based on Picard's iterations.\\
Let us recall the formulation of the Einstein-Vlasov-scalar field system. The spacetime is a four-dimensional manifold $M$, with local coordinates\\ $(x^\lambda) = (t, x^i )$ on which $x^0 = t$
denotes the time and $(x^i)$ the space coordinates. Greek indices always run from $0~ to~ 3$, and
Latin ones from $1~ to~ 3$. On $M$, a Lorentzian metric $g$ is given with signature $(-, +, +, +)$.
We consider a self-gravitating collision less gas and restrict ourselves to the case where all
particles have the same rest mass $m = 1$, and move forward in time. We denote by $(p^\lambda)$ the
momenta of the particles. The conservation of the quantity $g_{\lambda \beta }p^\lambda p^\beta$ requires that the phase
space of the particle is the seven-dimensional sub-manifold
\begin{equation*}
PM=\{g_{\mu
   \eta}p^{\mu}p^{\eta}=-1;~~p^0>0 \}
\end{equation*}
of $TM$ which is coordinatized by $(t,x^i,p^i)$. The energy-momentum tensor is given by
 \begin{equation}\label{1.2}
   T_{\mu\nu}\equiv  T^f_{\mu\nu}+ T^\phi_{\mu\nu}
\end{equation} with:
\begin{eqnarray}
  T^f_{\alpha\beta}&=&-\int_{\mathbb{R}^3}fp_{\alpha}p_{\beta}|g|^{1/2}\frac{dp^1dp^2dp^3}{p_0} \label{.Tf}\\
  T^\phi_{\alpha \beta} &=& \nabla_{\alpha}\phi\nabla_{\beta}\phi-\frac{1}{2}g_{\alpha
\beta}(\nabla_{\sigma}\phi\nabla^{\sigma}\phi+2V(\phi))\label{.Tfi}
\end{eqnarray}
where the distribution function of the particles is a non-negative real-valued function
denoted by $f$ and defined on $PM$,  \ $p_\lambda=g_{\lambda\beta} p^\beta,~ |g|$ denotes the modulus of determinant of
the metric $g$, a scalar field $\phi$ is a real-valued $C^\infty$ function on $M$, and $V $ is a positive $C^\infty$ real-valued function
(call potential)  
such that $V(0) = V_0 > 0,~ V'(0) = 0~ and~ V''(0) > 0$ (see [R]). \
  For $(t,~x^i,~p^i) \in PM$ \begin{equation*}
        p^0=\sqrt{-g^{00}}\sqrt{1+g_{ab}p^ap^b}
   \end{equation*}
 The Einstein field equations
   \begin{equation}\label{1}
    G_{\mu \nu}\equiv R_{\mu\nu}-\frac{1}{2}Rg_{\mu\nu}=8\pi T_{\mu\nu}
\end{equation}
should be coupled to the Vlasov equation (matter equation for $f$)
\begin{equation}\label{6}
    p^{\mu}\frac{\partial f}{\partial x^{\mu}}-\Gamma^{i}_{\nu
    \gamma}p^{\nu}p^{\gamma}\frac{\partial f}{\partial p^{i}}=0
 \end{equation}
 Since  $\nabla^{\alpha} G_\lambda \beta=0$ (Bianchi identities ) and the contribution of $f$ to the
energy-momentum tensor is divergence-free, ones obtain the following non-linear wave equation for $\phi$
 \begin{equation}\label{lc}
    \nabla^{\lambda}\nabla_{\lambda}\phi=V'(\phi).
    \end{equation}
\section{Equations in areal coordinates}
We refer to $[A]~or~ [BCIM]$ for details on the notion of $T^2$ symmetry. There are several
choices of spacetime manifolds compatible with $T^2$ symmetry. Here, we restrict our
attention to the $T^3$ case. Spacetimes admitting a $T^2$ isometry group acting on $T^3$ space-like
surfaces are more general than the $T^2$ spacetimes: both families admit two commuting
killing vectors. These spaces generalize Gowdy one. The dynamics of the matter is governed by the Vlasov
 and the non-linear wave equations. The
Vlasov equation models a collision-less system of particles which follow the geodesics of
space-time. We now consider a solution of the Einstein-Vlasov-scalar field system where all
unknowns are invariant under this symmetry. We write the system in areal coordinates, see $[A]$.
The circumstances under which coordinates of this type exist are discussed in $[A1]$ and
references therein. In such coordinates, the metric $g$ takes the form (we refer to
$([S],~[L],~[W]$)
\begin{equation}\label{2}
    ds^2=e^{2(\tau-\mu)}(-\alpha
    dt^2+d\theta^2)+e^{2\mu}[dx+Ady+(G+AH)d\theta]^2+e^{-2\mu}t^2[dy+Hd\theta]^2
\end{equation}
where $\mu,~\tau,~ \alpha,~G,~H,~ and~ A$ are unknown real functions of $t$ and $\theta$ variables, periodic in
$\theta$ with period $2 \pi.~$ Here, the timelike coordinate $t$ locally labels spatial hypersurfaces of the
spacetime, and each such hypersurface consists of all the $T^2$ group orbits with area $t$.
In order to simplify the equations, we introduce new quantities
\begin{equation}\label{2'}
    J =-\frac{t e^{-2\tau+4\mu}}{\sqrt{\alpha}}(G_t+AH_t),~~~~K=AJ-\frac{t^3
    e^{-2\tau}}{\sqrt{\alpha}}H_t.
\end{equation}
\begin{equation}\label{Gammat}
    \Gamma=G_t+AH_t
  \end{equation}
J and K are call twist quantities and both are zero in Gowdy symmetry case, .
 The scalar field $\phi$ is a function of $t$ and $\theta.~$  Using the result of $[S]$ or $[W]$, the complete
Einstein-Vlasov-scalar field system can be written in areal coordinates in the following form:
\subsubsection*{Vlasov equation}
\begin{eqnarray}
    ~&~&\left[\frac{\partial v_0}{\partial \theta}+\frac{\sqrt{\alpha}e^{2\tau}}{t^3}(K-AJ)(v_3-Av_2)
  +\frac{\sqrt{\alpha}e^{2\tau-4\mu}}{t}Jv_2 \right]\frac{\partial f}{\partial
  v_1} \nonumber \\
  ~&+&\left[\left(\frac{1}{t}-\mu_t\right)v^3-\sqrt{\alpha}\mu_{\theta}\frac{v^1v^3}{v^0}+e^{2\mu}\frac{v^2}{t}(A_t+\sqrt{\alpha}A_{\theta}\frac{v^1}{v^0})\right]\frac{\partial f}{\partial
  v_3} \nonumber \\
  ~&+&\left[\mu_tv^2+\sqrt{\alpha}\mu_\theta\frac{v^1v^2}{v^0}\right]\frac{\partial f}{\partial
  v_2}+\frac{\partial v_0}{\partial v_1}\frac{\partial f}{\partial
    \theta}=\frac{\partial f}{\partial t}  \label{Vl}
\end{eqnarray}
\subsubsection*{The Einstein constraint equations}
\begin{eqnarray}
  \frac{\tau_t}{t} &= \mu^2_t+\alpha \mu^2_\theta+\frac{e^{4\mu}}{4t^2}(A_t^2+\alpha A_{\theta}^2)+\frac{\alpha e^{2\tau-4\mu}}{4t^2}J^2 +\frac{\alpha e^{2\tau}(K-AJ)^2}{4t^4}\nonumber \\
  & +8\pi \frac{\sqrt{\alpha}}{t}\int_{\mathbb{R}^3} f|v_0|dv_1
  dv_2dv_3+\frac{1}{2}(\phi_t^2+\alpha\phi_{\theta}^2)+\alpha
  e^{2(\tau-\mu)}V(\phi) \label{16} \\
  \frac{\alpha_t}{\alpha} &= -\frac{\alpha
  e^{2\tau}}{t^3}(K-AJ)^2-2\alpha t e^{2(\tau-\mu)}V(\phi)
  \frac{-\alpha e^{2\tau-4\mu}}{t}J^2\nonumber \\
   & -16\pi
  \alpha^{\frac{3}{2}}e^{2(\tau-\mu)}\int_{\mathbb{R}^3}\frac{1+e^{-2\mu}v_2^2+e^{2\mu}t^{-2}(v_3-Av_2)^2}{|v_0|}f
  dv_1dv_2dv_3 \label{17} \\
 \frac{\tau_{\theta}}{t} &= 2\mu_t
 \mu_{\theta}+\frac{e^{4\mu}}{2t^2}A_tA_{\theta}-\frac{\alpha_{\theta}}{2t \alpha}-8\pi \frac{\sqrt{\alpha}}{t}\int_{\mathbb{R}^3}
 f v_1dv_1dv_2dv_3+ \phi_t \phi_{\theta} \label{18}
\end{eqnarray}
where $J~and~K$ are giving by $(\ref{2'})$.
\subsubsection*{The Einstein-matter evolution equations}
\begin{eqnarray}
  \tau_{tt}-\alpha \tau_{\theta \theta} &=& \frac{\alpha_{\theta }\tau_{\theta}}{2}+\frac{\tau_{t}\alpha_{t}}{2\alpha}-\frac{\alpha^2_{\theta}}{4\alpha}+\frac{\alpha_{\theta \theta}}{2}-\mu^2_t+\alpha \mu^2_{\theta}+\frac{e^{4\mu}}{4t^2}(A_t^2-\alpha A^2_{\theta})-\frac{\alpha e^{2\tau-4\mu} }{4t^2}J^2 \nonumber\\
  ~ &~& -3\frac{\alpha e^{2\tau}}{4t^4}(K-AJ)^2-8\pi\frac{\alpha^{\frac{3}{2}}e^{2\tau}}{t^3}\int_{\mathbb{R}^3}\frac{(v_3-Av_2)^2}{|v_0|}f
  dv_1dv_2dv_3 \nonumber \\
  ~ &~& -\frac{1}{2}(\phi_t^2-\alpha \phi_{\theta}^2)+\alpha e^{2(\tau-\mu)}V(\phi)  \label{19}
\end{eqnarray}
  \begin{eqnarray}
    \mu_{tt}-\alpha \mu_{\theta \theta} &=& \frac{-\mu_{t}}{t}+\frac{\alpha_{\theta}\mu_{\theta}}{2}+\frac{\alpha_t \mu_t}{2\alpha}+\frac{e^{4\mu}}{2t^2}(A^2_t-\alpha A^2_{\theta})+\frac{\alpha e^{2\tau-4\mu}}{2t^2}J^2 ~~~~~~~~~~~~~~~~~~~~~~~~~~\nonumber \\
    ~ &~& + 8\pi\frac{\alpha^{\frac{3}{2}}e^{2(\tau-\mu)}}{2t}\int_{\mathbb{R}^3}\frac{1+2e^{-2\mu}v_2^2}{|v_0|}fdv_1dv_2dv_3 \nonumber \\
    ~ &~& + 2\alpha e^{2(\tau-\mu)}V(\phi) \label{20}
  \end{eqnarray}
  \begin{eqnarray}
    A_{tt}-\alpha A_{\theta \theta} &=& \frac{A_t}{t}+\frac{\alpha_t A_t}{2\alpha}
    -4(A_t \mu_t-\alpha A_{\theta}\mu_{\theta})+\frac{\alpha e^{2\tau-4\mu}}{t^2}J(K-AJ) \nonumber  \\
    ~ &~&+\frac{\alpha_{\theta}A_{\theta}}{2} +16\pi \frac{\alpha^{\frac{3}{2}}e^{2(\tau-\mu)}}{t}\int_{\mathbb{R}^3}\frac{v_2(v_3-Av_2)}{|v_0|}fdv_1dv_2dv_3 \label{21}\\
  \phi_{tt}-\alpha \phi_{\theta \theta} &=&\left(\frac{\alpha_\theta}{2}+[2A_\theta H(G-1)-\mu_\theta(G+AH)^2]\alpha e^{-2(\tau-2\mu)}-4\mu_\theta
  H^2t^2\alpha e^{-2\tau}\right)\phi_\theta \nonumber\\
  ~ &+& \left(\frac{\alpha_t}{2\alpha}-\frac{1}{t}\right)\phi_t-\alpha e^{2(\tau-\mu)}V'(\phi)
 \label{22}
  \end{eqnarray}
\subsubsection*{The auxiliary equations}
\begin{eqnarray}
    \partial_\theta[e^{-2\tau}\alpha^{-1/2}e^{4\mu}\Gamma] &=& -2e^{\tau}\rho_{_2} \label{38ARW} \\
    \partial_t[e^{-2\tau}t\alpha^{-1/2}e^{4\mu}\Gamma] &=& 2t\alpha^{1/2}e^{\tau}S_{^{12}} \label{39ARW}\\
    \partial_\theta[e^{-2\tau}\alpha^{-1/2}(Ae^{4\mu}\Gamma+t^2H_t)]  &=&-2e^{\tau}A\rho_{_2}  -2te^{\tau-2\mu}\rho_{_3} \label{40ARW} \\
    \partial_t[e^{-2\tau}\alpha^{-1/2}(Ae^{4\mu}\Gamma+t^2H_t)] &=& 2t\alpha^{1/2}e^{\tau}(AS_{^{12}}+te^{-2\mu}S_{^{13}}). \label{41ARW}
  \end{eqnarray}
  where $\Gamma$ is defined by $(\ref{Gammat})$ and the matter terms defined by : ($k=2 \ or \ 3$)
  \begin{eqnarray}
  S_{1k} &=&-16 \pi \alpha \int_{\mathbb{R}^3}\frac{v_1v_k}{|v_0|}fdv_1dv_2dv_3 \label{23}\\
\rho_{_k} &=& 16 \pi \alpha \int_{\mathbb{R}^3}fv_kdv_1dv_2dv_3 \label{24}
  \end{eqnarray}
Since all the particles have proper mass $1$, the variables $v^\lambda$ are related to the canonical
momentum variables $p^\lambda$ by the relation
 \begin{eqnarray}
  (v^0)^2 &=& \alpha e^{2(\tau-\mu)}(p^0)^2 \label{cg1}\\
  (v^1)^2 &=&  e^{2(\tau-\mu)}(p^1)^2; \label{cg2}\\
  (v^2)^2 &=& e^{2\mu}[(G+AH)p^1+p^2+Ap^3]^2 \label{cg3}\\
  (v^3)^2 &=& t^2e^{-2\mu}(Hp^1+p^3)^2 \label{cg4}
\end{eqnarray}
\begin{equation*}
\text{so that} \  v^0=\sqrt{1+(v^1)^2+(v^2)^2+(v^3)^2}
\end{equation*}
 We prescribe initial data at time $t = t_0> 0 :$\\
 $~~~f(t_{_0},\theta,v)=f_{_0}(\theta,v)\equiv f^{^\circ} ,~\phi(t_{_0},\theta)=\phi_{_0}(\theta)\equiv\phi^{^\circ},~\phi_t(t_{_0},\theta)=\grave{\phi}(\theta)~$ and for all metric's component $\chi$, \ \
 $\chi(t_{_0},\theta)=\chi_{_0}(\theta)\equiv\chi^{^\circ}~and~\chi_t(t_0,\theta)=\grave{\chi}(\theta)$
\begin{definition}\label{defiregu}
  Let $I\subseteq ]0;+\infty[$ be an interval and $(t,\theta)\in I\times S^1$.
  \begin{enumerate}
    \item $f\in C^1(I\times S^1\times \mathbb{R}^2\times \mathbb{R}^3)$ is regular if $f(t,v,\theta+2\pi)=f(t,v,\theta)$ and
    supp$f(t,\theta,.,.,.,.,.)$ is uniformly compact in $\theta$ and locally uniformly in t .
    \item $\phi\in C^2(I\times S^1)$ is regular if $\phi(t,\theta+2\pi)=\phi(t,\theta)$ and $\phi(t,\theta)$  is uniformly compact in $\theta$ and locally uniformly in t.
\item  All component $\chi \in (I\times S^1)$ of metric is regular if $\chi(t,\theta+2\pi)=\chi(t,\theta)~~and \\ \partial_t\chi,~\partial_\theta \chi \in C^1(I\times S^1).$
\end{enumerate}
\end{definition}
\section{Estimations}
In this section, we proceed step by step to estimate all components of the metric, functions $f~and~\phi$ and their first derivatives.\\
Let us first prove an ''energy'' monotonicity lemma.
Let $E$ be defined by
\begin{equation}\label{E}
    E(t)= (E_{g}+E_{K}+E_{f}+E_{\phi})(t)
\end{equation}
where
\begin{equation}\label{Eg.}
    E_{g}(t)=\int_{S^1}\frac{1}{\sqrt{\alpha}}\left\{\mu^2_t+\alpha \mu^2_\theta+\frac{e^{4\mu}}{4t^2}(A_t^2+\alpha
 A_{\theta}^2)\right\}d\theta
\end{equation}
\begin{eqnarray}
  E_k(t)&=&\int_{S^1}\frac{1}{\sqrt{\alpha}}\left\{\frac{\alpha e^{2\tau-4\mu}}{4t^2}J^2 +\frac{\alpha e^{2\tau}(K-AJ)^2}{4t^4} \right\}d{\theta}\label{Ek.} \\
  E_f (t)&=&8\pi\int_{S^1} \left(\frac{1}{t}\int_{\mathbb{R}^3}f|v_0|dv_1dv_2dv_3\right)d\theta \label{Ef.}\\
E_\phi(t)&=&\int_{S^1}\left(\frac{1}{2}(\frac{1}{\sqrt{\alpha}}\phi_t^2+\sqrt{\alpha}\phi_{\theta}^2)+\sqrt{\alpha}
  e^{2(\tau-\mu)}V(\phi)\right)d\theta
   \label{E'}
 \end{eqnarray}
\begin{lemma} \label{lem Ef} $E$ is a monotonically decreasing function in t which satisfies for $t\geq t_{_0}$,
 \begin{eqnarray}
  \frac{dE}{dt} &=&- \int_{S^1} \left(\frac{2}{t}\frac{\mu^2_t}{\sqrt{\alpha}}+\frac{\phi_t^2}{t\sqrt{\alpha}}+\frac{e^{4\mu}}{2t^3}
  \sqrt{\alpha}A_{\theta}^2+\frac{3\sqrt{\alpha}}{4t^3}e^{2\tau-4\mu}J^2+\frac{\sqrt{\alpha} e^{2\tau}(K-AJ)^2}{t^5} \right)d\theta \nonumber\\
       ~ &-& \int_{S^1} \left[8\pi \int_{\mathbb{R}^3}\left(\frac{|v_0|}{t^2}+\frac{\alpha e^{2 \tau}(v_3-Av_2)^2}{t^4|v_0|}\right)f dv_1dv_2dv_3\right]d\theta \leq 0  \label{En}.
  \end{eqnarray}
\end{lemma}
\textbf{Proof}:
This is a straightforward but lengthy computation (see $[L]$ or $[W]$). Let us sketch the
steps involved. After taking the time derivative of the integrand we use the
evolution equations $(\ref{20})-(\ref{21})-(\ref{22})$ for $\mu,~A,~\phi$ to substitute for the second order derivatives, and the auxiliary equations to express second order derivatives of $G$ and
$H$ in terms of matter quantities. Integrating by parts and using the constraint equations for $\tau_t$ leads to $(\ref{En})$.\\
\textbf{step 1} : Bounds on $\phi$, $\alpha,~A,~\mu,\tau,~G,~H, ~K, ~J $.\\
Following step 1 in $[T2]$, we obtain uniform bound of $\phi$.
 About bounds of $\beta,~\mu,~A$ ($e^\beta=\sqrt{\alpha}e^\tau$), we use the same approach as in step 1 in $[A]$.
Since the function $t\longmapsto\sqrt{\alpha}$ is decreased, we deduce that $\alpha$ is bounded.\\ Now $\beta=\frac{1}{2}ln\alpha+\tau$
and $\alpha$ are bounded, then $\tau$ is bounded. Bounds on $e^{-\tau}G_t~ and~ e^{-\tau}H_t$  follow from   the end of step $1~of~[ARW]$. We deduce that $J~and~K$ are bounded. Consequently $H_t$, $G_t$, $H$ and $G$ are bounded.\\
\textbf{Step} $2$ \ Bounds of the first derivatives of $\tau,~A,~\mu,~\alpha, ~\phi$ and $f$. \\
To bound the derivatives of $\mu$ and $A$,  we use light-cone estimates in a similar way
as for the contracting direction (see. [LTN], [A], or [T2]). However, the matter terms must be treated
differently and we need to carry out a careful analysis of the characteristic
system associated with the Vlasov equation. Let us define
 \begin{equation}\label{74.}
    \Upsilon=\mu_t^2+\alpha\mu_{\theta}^2+\frac{e^{4\mu}}{4t^2}(A_t^2+\alpha
    A_{\theta}^2)
 \end{equation}
 \begin{eqnarray}
   \Upsilon^{\times} &=& 2\sqrt{\alpha}(\mu_t\mu_{\theta}+\frac{e^{4\mu}}{4t^2}A_t A_{\theta}) \label{74} \\
   \phi_{\varsigma} &=& \phi_t+\sqrt{\alpha}\phi_{\theta}  \label{74'}\\
   \phi_{\xi} &=& \phi_t-\sqrt{\alpha}\phi_{\theta}  \label{74''}
 \end{eqnarray}
 and the null paths
$\partial_\xi=\partial_t-\sqrt{\alpha}\partial_{\theta}~and~\partial_\varsigma=
 \partial_t+\sqrt{\alpha}\partial_{\theta}$, which induced that
$\phi^2_t+\alpha\phi^2_\theta=\frac{1}{2}(\phi^2_\xi+\phi^2_\varsigma)$.\\
Using the evolution equation $(\ref{20})$, a short computation shows that
\begin{eqnarray}
  \partial_\xi(\Upsilon+\Upsilon^{\times}) &=& -\frac{2}{t}(\mu_t^2+\frac{e^{4\mu}}{4t^2}\alpha A_{\theta}^2)+\frac{\alpha_t}{\alpha}(\Upsilon+\Upsilon^{\times})-\frac{\Upsilon^{\times}}{t} \nonumber \\
  ~ &~& +2(\mu_t+\sqrt{\alpha}\mu_{\theta})(\frac{e^{\beta-4\mu}}{2t^2}J^2 \nonumber \\
  ~ &~& +8\pi\frac{\sqrt{\alpha}e^{\beta-4\mu}}{2t}\int_{\mathbb{R}^3}\frac{1+2e^{-2\mu}v_2^2}{|v_0|}fdv_1dv_2dv_3
  + 2e^{2(\beta-\mu)}V(\phi)) \nonumber \\
  ~ &~& +\frac{e^{4\mu}}{4t^2}(A_t+\sqrt{\alpha}A_{\theta})[\frac{e^{\beta-4\mu}}{t^2}J(K-AJ)  \nonumber \\
  ~ &~&+16\pi \frac{\sqrt{\alpha}e^{2\beta-4\mu}}{2t}\int_{\mathbb{R}^3}\frac{v_2(v_3-Av_2)}{|v_0|}fdv_1dv_2dv_3]  \label{75}
\end{eqnarray}
and
\begin{eqnarray}
  \partial_\varsigma(\Upsilon-\Upsilon^{\times}) &=& -\frac{2}{t}(\mu_t^2+\frac{e^{4\mu}}{4t^2}\alpha A_{\theta}^2)+\frac{\alpha_t}{\alpha}(\Upsilon-\Upsilon^{\times})+\frac{\Upsilon^{\times}}{t} \nonumber \\
  ~ &~& +2(\mu_t-\sqrt{\alpha}\mu_{\theta})(\frac{e^{\beta-4\mu}}{2t^2}J^2 \nonumber \\
  ~ &~& +8\pi\frac{\sqrt{\alpha}e^{\beta-4\mu}}{2t}\int_{\mathbb{R}^3}\frac{f(1+2e^{-2\mu}v_2^2)}{|v_0|}dv_1dv_2dv_3+
  2e^{2(\beta-\mu)}V(\phi)) \nonumber \\
  ~ &~& +\frac{e^{4\mu}}{4t^2}(A_t-\sqrt{\alpha}A_{\theta})[\frac{e^{\beta-4\mu}}{t^2}J(K-AJ) \nonumber \\
  ~ &~&+16\pi \frac{\sqrt{\alpha}e^{2\beta-4\mu}}{2t}\int_{\mathbb{R}^3}\frac{fv_2(v_3-Av_2)}{|v_0|}dv_1dv_2dv_3]  \label{76}
\end{eqnarray}
Using $(\ref{17})$, we obtain:
\begin{eqnarray}
  |\partial_\xi(\Upsilon+\Upsilon^{\times})| &\leq& \left|\frac{\alpha_t}{\alpha}\right|\left(\frac{\Upsilon}{t}+\frac{1}{2t}+2\Upsilon\right)+\frac{3\Upsilon}{t}, \label{81} \\
  |\partial_\varsigma(\Upsilon-\Upsilon^{\times})| &\leq& \left|\frac{\alpha_t}{\alpha}\right|\left(\frac{\Upsilon}{t}+\frac{1}{2t}+2\Upsilon\right)+\frac{3\Upsilon}{t}, \label{82}.
\end{eqnarray}
Let
\begin{equation}\label{83}
  u_1 = \sqrt{\alpha}v_1
\end{equation}
\begin{eqnarray}
 \overline{u}_1(t) = sup\left\{|u_1|/ \exists(s,\theta,v_2,v_3) \in [t,t_i]\times S^1\times \mathbb{R}^2/f(s,\theta,v_1,v_2,v_3)\neq 0\right\} \label{84}\\
 \overline{v}_2(t) = sup\left\{|v_2|/ \exists(s,\theta,v_1,v_3) \in [t,t_i]\times S^1\times \mathbb{R}^2/f(s,\theta,v_1,v_2,v_3)\neq0\right\} \label{84+}\\
 \overline{v}_3(t) = sup\left\{|v_3|/ \exists(s,\theta,v_1,v_2) \in [t,t_i]\times S^1\times \mathbb{R}^2/f(s,\theta,v_1,v_2,v_3)\neq0\right\} \label{84++}
\end{eqnarray}
\begin{eqnarray}
\psi(t) &=& max\left(\underset{\theta \in S^1}{sup}\Upsilon(t,.)+\overline{u}^2_1(t);2\right)
  \label{85} \\
\Lambda(t) &=& \psi(t)+\chi(t),~~~\text{where}~~~\chi(t)=\underset{\theta\in
S^1}{sup}(|\phi_\xi|+|\phi_\varsigma|)(t,\theta)\label{85'+'} \\
  N(t,\theta) &=& \int_{t_0}^t\sqrt{\alpha}(s,\theta)ds \label{85'}.
\end{eqnarray}
Equation $(\ref{17})$ gives
\begin{equation*}
    -\frac{\alpha_t}{\alpha}=\frac{\alpha e^{2\tau-4\mu}}{t}J^2+\frac{\alpha
    e^{2\tau}}{t^3}(K-AJ)^2+\sqrt{\alpha}B
 \end{equation*}
 with \begin{equation}\label{86'}
    B(t,\theta)=16\pi
    e^{2\beta-2\mu}\int_{\mathbb{R}^3}\frac{1+e^{-2\mu}v_2^2+t^{-2}e^{2\mu}(v_3-Av_2)^2}{|v_0|}f dv_1dv_2dv_3
 \end{equation}
  By definition of $v_{_0}$, we have $ |v_{_0}|\geq\sqrt{\alpha e^{2(\tau-\mu)}+\alpha
 v_1^2}=e^{\beta-\mu}\sqrt{1+e^{-2\beta +2\mu}u_1^2}$. Then
 $(\ref{86'})$ becomes
 \begin{eqnarray}
   B(t,\theta) &\leq& 16\pi e^{2\beta-2\mu}\int_{\mathbb{R}^3}\frac{1+e^{-2\mu}v_2^2+t^{-2}e^{2\mu}(v_3-Av_2)^2}{e^{\beta-\mu}\sqrt{1+e^{-2\beta
   +2\mu}u_1^2}}fdv_1dv_2dv_3, \nonumber \\
   ~ &\leq & 16\pi e^{\beta-\mu}\|f_0\|\left(1+e^{-2\mu}(\bar{v}_2)^2+\frac{e^{2\mu}}{t^2}(\bar{v}_3+|A|\bar{v}_2)^2\right)4
   \bar{v}_2\bar{v}_3\int_{-\overline{u}_1}^{\overline{u}_1}\frac{du_1}{\sqrt{\alpha}\sqrt{1+e^{-2\beta+2\mu}u_1^2}} \nonumber \\
   ~ &\leq& 16\pi e^{\beta-\mu}\|f_0\|\left(1+e^{-2\mu}(\bar{v}_2)^2+\frac{e^{2\mu}}{t^2}(\bar{v}_3+|A|\bar{v}_2)^2\right)4
   \bar{v}_2\bar{v}_3 \nonumber \\
   ~ &~&
   \times2\left(e^{\beta-\mu}ln\left(\overline{u}_1+\sqrt{e^{2\beta-2\mu}+u_1^2}\right)+e^{-1}\right)
   \label{87}
   \end{eqnarray}
  We deduce that
\begin{eqnarray}
\frac{\alpha_t}{\alpha}&=&\frac{\alpha e^{2\tau-4\mu}}{t}J^2+\frac{\alpha e^{2\tau}}{t^3}(K-AJ)^2+\sqrt{\alpha}B \nonumber \\
    ~&\leq& c(t)+C(1+ln \overline{u}_1^2)\label{88}
 \end{eqnarray}
 We use similar approach as in inequality $(3.27)$ in $[T2]$ to show that
 \begin{eqnarray}
    \chi(t)&\leq&
    C+C(t)\int_{t_0}^t\left[\chi(s)(1+\bar{u}_1ln\bar{u}_1+ln\bar{u}_1(s))+\underset{\theta\in
    S^1}{sup}\Upsilon(s)(\chi(s))^2\right]ds\nonumber\\
    ~~&+&2C'\underset{\bar{t}\in\{t_0,t\}}{max}|V(\phi(\bar{t},\theta))|\label{85''}
\end{eqnarray}
where $\chi(t)$ is defined by
$(\ref{85'+'})~and~\Upsilon~by~(\ref{74.})$.
 We deduce from $(\ref{74.})$ that
\begin{eqnarray}
   \sqrt{\alpha}|\mu_{\theta}| &\leq& \sqrt{\Upsilon} \label{93}\\
    \sqrt{\alpha}\frac{e^{2\mu}|A_{\theta}|}{t} &\leq& 2\Upsilon+ \frac{1}{2} \sqrt{\Upsilon} \label{94}
 \end{eqnarray}
 Integrate  $(\ref{81})~and~(\ref{82})$  along the null paths $\partial_\xi,~\partial_\varsigma$ and add the two
 inequalities to obtain
 \begin{eqnarray}
  \Upsilon(t,\theta) &\leq&B+\int_{t_0}^t \left|\frac{\alpha_s}{\alpha}\right|\left((\frac{\Upsilon}{2s}+\frac{1}{2s}+\Upsilon)+\frac{3\Upsilon}{2s}\right)
  (s,\theta-N(s)+t_0)ds\nonumber  \\
  ~ &+&
  \int_{t_0}^t\left|\frac{\alpha_s}{\alpha}\right|\left((\frac{\Upsilon}{2s}+\frac{1}{2s}+g)+\frac{3\Upsilon}{2s}\right)
  (s,\theta+N(s)-t_0)ds
  \label{99}
  \end{eqnarray}
 Adding $(\ref{85''})-(\ref{99})$ to obtain 
\begin{equation}\label{103}
\Lambda(t)\leq C+C(t)\int_{[t_0;t]}(\Lambda(s))^3ln\Lambda(s)ds
\end{equation}
where $\Lambda(t)$ is defined by $(\ref{85'+'})$. Using Gronwall inequality, we conclude from $(\ref{103})$ that
$\Lambda(t)$ is uniformly bounded for any $t\in [t_0,t_1)$. And consequently,  first derivatives of $\mu,~A,~\phi$ are bounded. The bounds of $\alpha_t$ and $\tau_t$ follows from $(\ref{17})$ and $(\ref{18})$ . Uniform bounds of $\alpha_\theta$ and $\tau_\theta$ follow from step 3 in [T2]. Uniform bounds of $\partial f$ follow from step 4 in $[T2]$.\\
\section{Local Existence}
In this section, we prove using an iteration the local in time existence and uniqueness of solutions of the Einstein-Vlasov-Scalar field system. We denote by $\nu:=\nu(t;\theta)~~for~all~(t,\theta)\in]t_0;T)\times S^1$, all component of the metric; and $\nu_\theta:=\nu'~and~\nu_t:=\dot{\nu}$.\\
If $\alpha_{n-1},~\tau_{n-1},~\mu_{n-1},~A_{n-1},~\alpha'_{n-1},~\tau'_{n-1},~\mu'_{n-1},~A'_{n-1}$ are already defined and regular on $]t_0;T)\times S^1$ then let
\begin{eqnarray}
  \Xi_{n-1}(s,t,\theta,V) &= \left[\frac{V^1\sqrt{\alpha_{n-1}}}{V^0}; \Xi_{n-1}^1; \Xi_{n-1}^2; \Xi_{n-1}^3\right]~~\label{4.2} \\
 \text{where}~~~\Xi_{n-1}^1=&-(\tau'_{n-1}-\mu'_{n-1}+\frac{\alpha'_{n-1}}{2\alpha_{n-1}})\sqrt{\alpha_{n-1}}V^0-(\dot{\tau}_{n-1}
 -\dot{\mu}_{n-1})V^1 \nonumber \\
 &-\frac{\sqrt{\alpha_{n-1}}\mu'_{n-1}}{V^0}((V^2)^2-(V^3)^2)+\frac{\sqrt{\alpha_{n-1}}
A'_{n-1}}{sV^0}e^{2\mu_{n-1}}V^2V^3; \nonumber \\
 \Xi_{n-1}^2=&-\dot{\mu}_{n-1}
V^2-\sqrt{\alpha_{n-1}}\mu'_{n-1}\frac{V^1V^2}{V^0} ;\nonumber\\
\Xi_{n-1}^3=&-\left(\frac{1}{s}-\dot{\mu}_{n-1}\right)V^3
-\sqrt{\alpha_{n-1}}\mu'_{n-1}\frac{V^1V^3}{V^0}-\frac{e^{2\mu_{n-1}}}{s}\left(\dot{A}_{n-1}
+\sqrt{\alpha_{n-1}}A'_{n-1}\frac{V^1}{V^0}\right)V^2\nonumber
\end{eqnarray}
and denote by $(\Theta_n;V_n)(s,t,\theta,V^1,V^2,V^3)$ the solution of the characteristic system
\begin{equation*}
    \frac{d}{ds}\left(\Theta_n,V_n\right)=\Xi_{n-1}(s,t,\Theta,V)
\end{equation*}
with initial data $(\Theta_n,V_n)(t_0,t,\theta,V^1,V^2,V^3)=(\theta,v)$. Define
\begin{equation}\label{2+}
    f_n(t,\theta,V^1,V^2,V^3)=f_{_0}((\Theta_n,V_n)(t_{_0},\theta,V^1,V^2,V^3))
\end{equation}
where $f_n$ is solution of
 \begin{eqnarray}
    ~&~&\left[\frac{\partial v_0}{\partial \theta}+\frac{\sqrt{\alpha_{_{n-1}}}e^{2\tau_{_{n-1}}}}{t^3}(K_{_{n-1}}-A_{_{n-1}}J_{_{n-1}})(v_3-A_{n-1}v_2)
  +\frac{\sqrt{\alpha_{_{n-1}}}e^{2\tau_{_{n-1}}-4\mu_{_{n-1}}}}{t}J_{_{n-1}}v_2 \right]\frac{\partial f_n}{\partial
  v_1} \nonumber \\
  ~&+&\left[\left(\frac{1}{t}-\dot{\mu}_{_{n-1}}\right)v^3-\sqrt{\alpha_{_{n-1}}}\mu'_{n-1}
  \frac{v^1v^3}{v^0}+e^{2\mu_{^{n-1}}}\frac{v^2}{t}(\dot{A}_{_{n-1}}+\sqrt{\alpha_{_{n-1}}}
  A'_{_{n-1}}\frac{v^1}{v^0})\right]\frac{\partial f_n}{\partial
  v_3} \nonumber \\
  ~&+&\left[\dot{\mu}_{_{n-1}}v^2+\sqrt{\alpha_{_{n-1}}}\mu'_{_{n-1}}\frac{v^1v^2}{v^0}\right]\frac{\partial f_n}{\partial
  v_2}+\frac{\partial v_0}{\partial v_1}\frac{\partial f_n}{\partial
    \theta}=\frac{\partial f_n}{\partial t}  \label{4.3}
\end{eqnarray}
with $f_{_n}(t_{_0})=f_{_0}$.\\
Define $\rho_{_n},~\rho_{_{k,n}},~P_{_{k,n}},~k\in\{1;2;3\}$  using the formulas $\rho,~\rho_{_k},~P_k$.  Respectively replaced
\begin{equation*}
f,~\alpha,~\dot{\phi},~\phi',~\dot{A},~A',~\mu,~\tau,~\dot{\mu},~\dot{\tau},~\mu',~\tau'
\end{equation*}
\begin{equation*}
\text{by} \  f_{_n},~\alpha_{_{n-1}},~\dot{\phi}_{_{n-1}},~\phi'_{_{n-1}},~\dot{A}_{_{n-1}},~A'_{_{n-1}},~\mu_{_{n-1}},
~\tau_{_{n-1}},~\dot{\mu}_{_{n-1}},~\dot{\tau}_{_{n-1}},~\mu'_{_{n-1}},~\tau'_{_{n-1}}.
\end{equation*}
$~~$ Use formulas
$(\ref{74.}),~(\ref{74}),~(\ref{74'})~et~(\ref{74''})~$ to define respectively
$\Upsilon_n,~~\Upsilon^\times_n,~~\phi_{\varsigma_n}~~and~~\phi_{\xi_n}$.
 Let
\begin{eqnarray}
 \bar{\Upsilon}_n&=& \Upsilon_n+\dot{\phi}_n^2+\alpha_n\phi'^2_n \label{4.3'} \\
  \bar{\Upsilon}^\times_n &=&
  \Upsilon^\times_n+2\sqrt{\alpha_n}\dot{\phi}_n \phi'_n. \label{4.3''}
\end{eqnarray}
Using constraint equations,  we define
 $\alpha_n,~A_n,~J_n,~K_n,~\tau_n,~\mu_n,~and~\phi_n~$ solutions of the system
 \begin{eqnarray}
  &\dot{\tau}_n = t\Upsilon_n+\frac{\alpha_n e^{2\tau_n-4\mu_n}}{4t}J_n^2 +\frac{\alpha_n e^{2\tau_n}(K_n-A_nJ_n)^2}{4t^3}+\alpha_ne^{2(\tau_n-\mu_n)}\rho_n \label{4.5} \\
  &\frac{\dot{\alpha}_n}{\alpha_n} = \frac{-\alpha_n e^{2\tau_n-4\mu_n}}{t}J_n^2-\alpha_n\frac{e^{2\tau_n}(K_n-A_nJ_n)^2}{t^3}+\sqrt{\alpha_n}(\rho_{_n}-P_{1;n}) \label{4.6} \\
&\tau'_{n}= t\Upsilon^{\times}_n-\frac{\alpha'_n}{2\alpha_n}-
\sqrt{\alpha_n}\rho_{_{1,n}}\label{4.7}
\end{eqnarray}
In the same way,  replace all component of the metric $\nu~by~ \nu_{n-1}$ in
 $(\ref{22})$ to obtain
 \begin{eqnarray}
  \partial_{\xi_{n-1}}
  \phi_{\varsigma_n}&=&\left(\frac{\dot{\alpha}_{n-1}}{2\alpha_{n-1}}-\frac{1}{2t}+\daleth_{n-1}\right)
  \phi_{\varsigma_{n-1}}\nonumber\\
  ~&~&-\left(\frac{1}{2t}+\daleth_{n-1}\right)\phi_{\xi_{n-1}}-e^{2\beta_{n-1}-2\mu_{n-1}}V'(\phi_{n-1}) \label{4.8}\\
  \partial_{\varsigma_{n-1}}\phi_{\xi_{n}}&=&\left(\frac{\dot{\alpha}_{n-1}}{2\alpha_{n-1}}-\frac{1}{2t}-
  \daleth_{n-1}\right)\phi_{\xi_{n-1}}\nonumber \\
  ~&~&-\left(\frac{1}{2t}-\daleth_{n-1}\right)\phi_{\varsigma_{n-1}}-e^{2\beta_{n-1}-2\mu_{n-1}}V'(\phi_{n-1}) \label{4.9}
 \end{eqnarray}
 here
 \begin{eqnarray}
  \daleth_{n-1}&=&\alpha_{n-1}(2A'_{n-1} H_{n-1}(G_{n-1}-1)-2\mu'_{n-1}(G_{n-1}+A_{n-1}H_{n-1})^2e^{-2(\tau_{n-1}-2\mu_{n-1})}\nonumber \\
  ~&~&-4\mu'_{n-1}H_{n-1}^2t^2e^{-2\tau_{n-1}}) \label{4.10}
  \end{eqnarray} \\
Let $\tilde{\Upsilon}_n~and~\breve{\Upsilon}_n$ be as follow
\begin{equation}\label{11}
    \tilde{\Upsilon}_n =\Upsilon_n+\Upsilon^{\times}_n,~~\breve{\Upsilon}_n =\Upsilon_n-\Upsilon^{\times}_n
\end{equation}
Then
\begin{eqnarray}
\partial_{\xi_{n-1}}\tilde{\Upsilon}_n
  &=(\frac{\dot{\alpha}_{n-1}}{\alpha_{n-1}}-\frac{1}{2t})\tilde{\Upsilon}_{n-1}-\frac{1}{2t}
  \breve{\Upsilon}_{n-1}+ a^+_{n-1} \label{4.12}\\
\partial_{\varsigma_{n-1}}\breve{\Upsilon}_n
  &=\frac{1}{2t}\tilde{\Upsilon}_{n-1}+(\frac{\dot{\alpha}_{n-1}}{\alpha_{n-1}}-\frac{1}{2t})
  \breve{\Upsilon}_{n-1}+ a^{-}_{n-1} \label{4.13}
\end{eqnarray}
where
\begin{eqnarray}
a^{+}_{n-1} &=& \frac{e^{4\mu_{n-1}}}{4t^2}A_{_{\varsigma_{_{n-_1}}}}\left[\frac{e^{2\beta_{n-1}-4\mu_{n-1}}}{t^2}
 J_{n-1}(K_{n-1}-A_{n-1}J_{n-1})+2 \sqrt{\alpha_{n-1}}\bar{S}_{23,n-1}\right] \nonumber \\
  ~ &~& +2\mu_{\varsigma_{n-_1}}\left[\frac{e^{\beta_{n-1}-4\mu_{n-1}}}{2t^2}J_{n-1}^2+\sqrt{\alpha_{n-1}}
  (\tilde{\bar{\rho}}_{n-1}-\tilde{\bar{P}}_{1,{n-1}}+\tilde{\bar{P}}_{2,{n-1}}-\bar{P}_{3,{n-1}})\right]\nonumber \\
  ~ &-&\frac{2}{t}(\dot{\mu}_{n-1}^2+\frac{e^{4\mu_{n-1}}}{4t^2}\alpha_{n-1} A^{'2}_{n-1})+\frac{\dot{\alpha}_{n-1}}{\alpha_{n-1}} \label{4.14}
\end{eqnarray}
 The expression of $a^-_{n-1}$ is written as $a^+_{n-1}$ replacing $\varsigma_{_{n-1}}~by~\xi_{_{n-1}}$.\\
 With these notations,  $(\ref{38ARW})$ and $(\ref{40ARW}$ give 
\begin{eqnarray}
  \partial_\theta J_{n+1} &=& -2te^{\beta_n}\rho_{_{2,n}}  \label{39ARWn} \\
  \partial_\theta K_{n+1}&=& -2t^{-1}e^{\beta_n}A_n\rho_{_{2,n}}-2t^{-2}e^{\tau_n -2\mu_n}\rho_{_{3,n}}      \label{41ARWn}
\end{eqnarray}.\\
We introduce the following quantities which are similar to those defined in
$(\ref{83})-(\ref{85'})$
\begin{eqnarray}
 & u_{n;1} = \sqrt{\alpha_n}v_1 \label{4.14}\\
 &\overline{u}_{n;1}(t) = sup\left\{u_{n;1}/ \exists(s,\theta,v_2,v_3) \in [t,t_i]\times S^1\times \mathbb{R}^3/f_n(s,\theta,v)\neq 0\right\} \label{4.15}\\
 &\psi_n(t) = max\left(\underset{\theta \in S^1}{sup}\Upsilon_n(t,.)+\overline{u}^2_{n;1}(t);2\right)
  \label{4.16} \\
 &\Lambda_n(t)= \psi_n(t)+\chi_n(t),~~\text{where}~~\chi_n(t):=\underset{\theta
  \in S^1}{sup}(\phi_{\xi_n}+\phi_{\varsigma_n})(t)\label{4.17}\\
 &D_n(t)= \sup\{\|\partial_\theta f_n(s)\|;t_0\leq s \leq t, for~all~t\in[t_0;T]\}
  \label{4.17n}\\
 &N_n(t,\theta)= \int_{t_0}^t\sqrt{\alpha_n}(s,\theta)ds \label{4.18}.
\end{eqnarray}
We follow the work of $[JLS]$ to prove that iterates are bounded and converge.
\begin{theorem}\label{theoEn} Consider initial data $\mu^{^\circ},~\tau^{^\circ},~\alpha^{^\circ},~A^{^\circ},~\phi^{^\circ},~K^{^\circ},~J^{^\circ}$ as well as $\grave{\mu},~\grave{\tau},~\grave{A},~\grave{\phi}$ satisfying   $(\ref{4.5})-(\ref{4.7})$, then
\begin{equation}\label{4.19}
    \mathcal{E}_n(t)\lesssim \mathcal{E}^{^\circ}
\end{equation}
with
 \begin{eqnarray}
\mathcal{E}_n(t)&=&\int_{S^1}\frac{1}{\sqrt{\alpha_n}}\left\{\Upsilon_n+\frac{\alpha_n e^{2\tau_n-4\mu_n}}{4t^2}J_n^2 +\frac{\alpha_n e^{2\tau_n}(K_n-A_nJ_n)^2}{4t^4} \right\}d{\theta}\nonumber \\
  ~ &~& +8\pi\int_{S^1} \left(\frac{1}{t}\int_{\mathbb{R}^3} f_n|v_0|dv_1
  dv_2dv_3\right)d\theta  \nonumber \\
  &~&+\int_{S^1}\left(\frac{1}{2}(\frac{1}{\sqrt{\alpha_n}}\dot{\phi}_n^2+\sqrt{\alpha_n}{\phi'}_n^2)+\sqrt{\alpha_n}
  e^{2(\tau_n-\mu_n)}V(\phi_n)\right)d\theta. \label{4.20} \\
  \mathcal{E}^{^\circ}&=&\int_{S^1}\frac{1}{\sqrt{\alpha^{^\circ}}}\left\{(\dot{\mu}^{^\circ})^2+
  \sqrt{\alpha^{^\circ}}\grave{\mu}'^2+\frac{e^{4\mu^{^\circ}}}{4t_{_0}^2}(\dot{A}^{{^\circ}^2}+
  \sqrt{\alpha^{^\circ}}\grave{A}'^2)\right\} d{\theta}\nonumber \\
 ~~&+&\int_{S^1}\left(\frac{1}{2}(\frac{1}{\sqrt{\alpha^{^\circ}}}(\dot{\phi}^{^\circ})^2+
   \sqrt{\alpha^{^\circ}}\grave{\phi}'^2)+\sqrt{\alpha^{^\circ}}e^{2(\tau^{^\circ}-\mu^{^\circ})}V(\phi^{^\circ})
   +\frac{\alpha^{^\circ} e^{2\tau^{^\circ}-4\mu^{^\circ}}}{4t_{_0}^2}J^{^{\circ^2}}\right)d\theta \nonumber\\
    ~&+&8\pi\int_{S^1} \left(\frac{1}{t_0}\int_{\mathbb{R}^3} f^{^\circ}|v_0|dv_1dv_2dv_3+\frac{\alpha^{^\circ} e^{2\tau^{^\circ}}(K^{^\circ}-A^{^\circ}J^{^\circ})^2}{32\pi t_0^4}\right)d\theta. \label{4.21}
\end{eqnarray}
\end{theorem}
\textbf{Proof}:
We use similar approach like in the proof of lemma $\ref{lem Ef}$ to obtain
\begin{eqnarray}
 \frac{d\mathcal{E}_{_n}}{dt} &=&- \int_{S^1} \left(\frac{2}{t}\frac{\dot{\mu}^2_n}{\sqrt{\alpha_n}}+\frac{\dot{\phi}_n^2}{t\sqrt{\alpha_n}}+
 \frac{e^{4\mu_n}}{2t^2}\sqrt{\alpha_n}{A'}_{_n}^2+\frac{3\sqrt{\alpha_n}}{4t^3}e^{2\tau_n-4\mu_n}J_n^2+
 \frac{\sqrt{\alpha_n} e^{2\tau_n}(K_n-A_nJ_n)^2}{t^4} \right)d\theta \nonumber\\
 ~ &-& \int_{S^1} \left[8\pi \int_{\mathbb{R}^3}\left(\frac{f_n|v_0|}{t^2}+\frac{\alpha_ne^{2 \tau}f(v_3-A_nv_2)^2}{t^4|v_0|}\right)dv_1dv_2dv_3\right]d\theta \leq 0  \label{4.24};
\end{eqnarray}
and since $\mathcal{E}_n$ is decreasing in time, for all $t\in [t_{_0},t_1],$~  $(\ref{4.19})$ follows.\\
 The next five lemmas and  proposition prove the following assertions :
 \begin{enumerate}
\item uniform bound on the momenta in the support of distribution function $f_n$,
\item uniform bound of the first derivatives with respect to  $\theta$ of $\mu_n,~A_n,~\tau_n,~\alpha_n,~\phi_n,~K_n~and~J_n $,
\item convergence of the iterates.
\end{enumerate}
\begin{lemma} \label{lem4.1}
$\inf\limits_{\theta\in S^1}\alpha_n(t;.)$ is uniformly bounded on $(t_{_0};t_{_1}]$.
\end{lemma}
\textbf{Proof}:
Using the expression of  $\mathcal{E}_n$, we have
\begin{equation}\label{4.25}
8\pi\int_{S^1}\left(\int_{\mathbb{R}^3}f_n|v^0|dv^1dv^2dv^3\right)d\theta\leq
t\mathcal{E}_n
\end{equation}
Since  $\sqrt{\alpha_n}|v_1|\leq|v^0|$,  we deduce progressively that
\begin{eqnarray}
\int_{S^1}\left(\int_{\mathbb{R}^3}f_n\sqrt{\alpha_n}|v_1|dv_1dv_2dv_3\right)d\theta&\leq& \frac{t\mathcal{E}_n}{8\pi}\lesssim\frac{t\mathcal{E}^{^\circ}}{8\pi} \nonumber  \\
\sqrt{\inf\limits_{\theta\in S^1}\alpha_n}\int_{S^1}\left(\int_{\mathbb{R}^3}f_n|v_1|dv^1dv^2dv^3\right)d\theta &\le&  \frac{t\mathcal{E}_n}{8\pi}\lesssim\frac{t\mathcal{E}^{^\circ}}{8\pi} \nonumber \\
~~  \sqrt{\inf\limits_{\theta\in S^1}\alpha_n} &\leq&
  \frac{t\mathcal{E}_n}{8\pi\delta}\lesssim \frac{t\mathcal{E}^{^\circ}}{8\pi\delta}   \label{4.26}
\end{eqnarray}
And the proof is completed.\\
Let us define the following expressions
\begin{equation}\label{4.27}
\beta_n =\tau_n +\frac{\ln\alpha_n}{2},~~\forall n \in \mathbb{N}^{*}
\end{equation}
 Taking partial derivative with respect to $\theta$ of $(\ref{4.27})$  and using $(\ref{4.7})$ give
\begin{equation}\label{4.28}
{\beta'}_n=\frac{2t}{\sqrt{\alpha}}\Upsilon^\times_n-\sqrt{\alpha}\left(\int_{\mathbb{R}^3}f_n|v_1|dv^1dv^2dv^3\right)
 \end{equation}
\begin{lemma}\label{lem4.2}  $\int_{S^1}|{\beta'}_n|d\theta,~\int_{S^1}|{\mu'}_n|d\theta,~\int_{S^1}|{\phi'}_n|d\theta,
~\int_{S^1}|{K'}_n|d\theta,~\int_{S^1}|{J'}_n|d\theta$
are bounded.
\end{lemma}
\textbf{Proof}:
$\alpha_n(t)~and~\mathcal{E}_n(t)$ are decreased in time
and using definition of $\mathcal{E}_n$,
\begin{eqnarray}
  \int_{S^1}|{\beta'}_n|d\theta &\leq&t\mathcal{E}_n\lesssim t\mathcal{E}^{^\circ} \label{4.29}\\
  \int_{S^1}|{\mu'}_n|d\theta &\leq& \frac{\sqrt{\mathcal{E}_n}}{\alpha_n}\lesssim\frac{\sqrt{\mathcal{E}^{^\circ}}}{(\inf\limits_{S^1}\alpha_n)^{1/4}} \label{4.30}\\
   \int_{S^1}|e^{2\mu_n}A'_n|d\theta &\leq& \frac{\sqrt{2t\mathcal{E}_n}}{(\inf\limits_{S^1}\alpha_n)^{1/4}}\lesssim \frac{\sqrt{2t\mathcal{E}^{^\circ}}}{(\inf\limits_{S^1}\alpha_n)^{1/4}} \label{4.31}\\
   \int_{S^1}|{\phi'}_n|d\theta &\leq& \frac{\sqrt{2\mathcal{E}_n}}{(\inf\limits_{S^1}\alpha_n)^{1/4}}
   \lesssim\frac{\sqrt{2\mathcal{E}^{^\circ}}}{(\inf\limits_{S^1}\alpha_n)^{1/4}}
   \label{4.32}\\
   (\ref{39ARWn})\Rightarrow \int_{S^1}|J'_n(t,\theta)|d\theta &\le & C\bar{v_2}\label{4.32n}\\
   (\ref{41ARWn})\Rightarrow \int_{S^1}|K'_n(t;\theta)|d\theta &\le & C(\bar{v_2}+ \bar{v_3})\label{4.32nn}
\end{eqnarray}
\begin{lemma}\label{lem4.3} For all $r,~b\in \mathbb{R},~~e^{r\beta_n+b\mu_n}A_n$ is bounded on $(t_{_0},t_1]\times S^1$.
\end{lemma}
\textbf{Proof}:
  Let $\bar{\theta}\in S^1$. \ \
 $\alpha_n$ is bounded on  $(t_0,t_1]\times \{\bar{\theta}\},~(see~(\ref{4.26}))$  then $e^{r\beta_n+b\mu_n}A_n$  is bounded on $(t_0,t_1]\times
 \{\bar{\theta}\}$. Integrating from
 $(t;\bar{\theta}) ~to~(t;\hat{\theta})~with \\ \bar{\theta}<\hat{\theta}$
  gives :
 \begin{eqnarray*}
   e^{r\beta_n+b\mu_n}A_n(t,\hat{\theta}) &=& (e^{r\beta_n+b\mu_n}A_n)(t,\bar{\theta})\\
   &+& \int_{\bar{\theta}}^{\hat{\theta}}
   \left(e^{r\beta_n+b\mu_n}A_n(r{\beta}'_n+b{\mu}'_n)
 + e^{r\beta_n+(b-2)\mu_n}e^{2\mu_n}{A'}_n\right)d\theta\\
   |e^{r\beta_n+b\mu_n}A_n(t,\hat{\theta})| &\leq& \left|(e^{r\beta_n+b\mu_n}A_n)(t,\bar{\theta})\right|\\
   &+& \left|\int_{\bar{\theta}}^{\hat{\theta}}\left(e^{r\beta_n+b\mu_n}A_n(r{\beta}'_n+b{\mu}'_n) +e^{r\beta_n+(b-2)\mu_n}e^{2\mu_n}{A'}_n\right)d\theta\right|\\
   ~~ &\leq&C+ \int_{\bar{\theta}}^{\hat{\theta}}\left(e^{r\beta_n+b\mu_n}|A_n|(|r{\beta}'_n|+|b{\mu}'_n|) \right)d\theta
\end{eqnarray*}
where $C$ depends on 
$|(e^{r\beta_n+b\mu_n}A_n)(t,\bar{\theta})|$ and the bounds obtain in lemma $\ref{lem4.2}$. Using Gronwall's inequality, one obtains:
\begin{equation}\label{4.33}
|e^{r\beta_n+b\mu_n}A_n(t,\hat{\theta})| \leq C
e^{|\int_{\bar{\theta}}^{\hat{\theta}}(r{\beta}'_n+b{\mu}'_n)d\theta|}
\end{equation}
 \begin{lemma} \label{lem4.4} For all $t\in(t_0,t_1]$,
 \begin{equation*}
   \int_{t_0}^{t}\sup\limits_{S^1}\left[e^{2\beta_n-4\mu_n}J^2_n\right](s,\theta)ds
 \end{equation*}
is bounded on $(t_0,t_1]$.
 \end{lemma}
\textbf{Proof}: similar to the proof of lemma $\ref{lem4.3}$. 
 We have this inequality
 \begin{equation*} 
   \int_{t_0}^{t}\left[e^{2\beta_n-4\mu_n}J^2_n\right](s,\theta)ds \leq  \int_{t_0}^{t}e^C\left[e^{2\beta_n-4\mu_n}J^2_n+2C|J_n|+C^2\right](s,\bar{\theta})ds
    \end{equation*}
which comes from
    \begin{eqnarray}
\int_{t_0}^{t}\sup\limits_{S^1}\left[e^{2\beta_n-4\mu_n}J^2_n\right](s,\theta)ds &\leq& \int_{t_0}^{t}\sup\limits_{S^1}\int_{\bar{\theta}}^{\theta}\left[e^{2\beta_n-4\mu_n}
J^2_n(2\beta'_n-4\mu'_n)\right](s,\check{\theta})d\check{\theta}ds \nonumber\\
~ &+&\int_{t_0}^{t}\sup\limits_{S^1}\left[e^{2\beta_n-4\mu_n}J^2_n\right](s,\bar{\theta})ds
      \label{4.35}
    \end{eqnarray}
The proof is completed applying Gronwall inequality to $(\ref{4.35})$.
\begin{lemma} \label{lem4.5} For all $t\in(t_0,t_1]$, the quantity
\begin{equation*}  
  \int_{t_0}^{t}\sup\limits_{S^1}\left[e^{2\beta_n(s,\theta)}(K_n-A_nJ_n)^2\right](s,\theta)ds
\end{equation*}
is bounded on $(t_0,t_1].$
\end{lemma}
 \textbf{Proof}: Choose $~\bar{\theta},~\hat{\theta}\in
S^1$ such as $\bar{\theta}<\hat{\theta}$.  Integrate $~[e^{\beta_n}(K_n-A_nJ_n)]'$  with respect to
$\theta$, from $(t,\bar{\theta})$ to  $(t,\hat{\theta})$, to obtain
\begin{eqnarray*}
&|e^{\beta_n}(K_n-A_nJ_n)|(t,\hat{\theta}) \leq [e^{\beta_n}|K_n-A_nJ_n|](t,\bar{\theta})+|\int_{\bar{\theta}}^{\hat{\theta}}[e^{\beta_n}|K_n-A_nJ_n| \nonumber\\
&+ e^{\beta_n}\int_{\mathbb{R}^3}f_n|v_2|dv+ e^{\beta_n}A_n\int_{\mathbb{R}^3}f_n|v_2|dv
+e^{\beta_n-2\mu_n}|J_n|e^{2\mu_n}|A'_n|](t,\check{\theta})d{\check{\theta}}|\nonumber \\
&\leq[e^{\beta_n}|K_n-A_nJ_n|](t,\bar{\theta})+C'
+C\sup\limits_{S^1}(e^{\beta_n-2\mu_n}|J_n|+\int_{\bar{\theta}}^{\hat{\theta}}e^{\beta_n}|K_n-A_nJ_n|
|{\beta'}_n|d\check{\theta}) 
  \end{eqnarray*}
  here
  $C'=[\int_{S^1}\left(\int_{\mathbb{R}^3}f_ndv\right)d\theta]sup(v_2e^{\beta_n}+v_3A_ne^{\beta_n})$.
  Using Gronwall's inequality, one have
  \begin{eqnarray}
    |e^{\beta_n}(K_n-A_nJ_n)|(t,\hat{\theta}) &\leq&e^{\int_{\bar{\theta}}^{\theta}|\beta'_n|d\theta}[(e^{\beta_n}(K_n-A_nJ_n)(t,\bar{\theta} \nonumber\\
    ~ &+&C'+C\sup\limits_{S^1}(e^{\beta_n-2\mu_n}|J_n|)] \nonumber\\
\Rightarrow\sup\limits_{S^1}C[(K_n-A_nJ_n)](t,\hat{\theta}) &\leq& e^{\int_{\bar{\theta}}^{\theta}|\beta'_n|d\theta}[(e^{\beta_n}(K_n-A_nJ_n)(t,\bar{\theta}) \nonumber\\
~ &+&C'+C\sup\limits_{S^1}(e^{\beta_n-2\mu_n}|J_n|)) \label{4.38}
  \end{eqnarray}
  therefore
  \begin{eqnarray}
  &2e^{\beta_n}|K_n-A_nJ_n|(s,\check{\theta})C\sup\limits_{S^1}e^{\beta_n-2\mu_n}|J_n| \nonumber\\
  &\leq e^{2\beta_n}(K_n-A_nJ_n)^2(s,\check{\theta})
    +C^2\sup\limits_{S^1}e^{2\beta_n-4\mu_n}J^2_n(s,\check{\theta})  \label{4.39}
  \end{eqnarray}
  In other ways
  \begin{eqnarray}
    &\int_t^{t_i}\sup\limits_{S^1}(e^{\beta_n}(K_n-A_nJ_n)^2)(s,\bar{\theta}))ds \leq e^{2C}\int_t^{t_i}e^{2\beta_n(s,\hat{\theta})}(K_n-A_nJ_n)^2(s,\hat{\theta})ds \nonumber\\
  &+ 2C'(e^{\beta_n}|K_n-A_nJ_n|)(s,\hat{\theta})+C'+C \sup\limits_{S^1}(e^{(\beta_n-2\mu_n})|J_n|)^2\nonumber\\
  &+2(e^{\beta_n}|K_n-A_nJ_n|)(s,\hat{\theta})C\sup\limits_{S^1}(e^{\beta_n-2\mu_n}|J_n|) \label{4.40}
  \end{eqnarray}
   Gronwall's lemma show again that
  \begin{equation}
\int_t^{t_i}\sup\limits_{S^1}(e^{\beta_n}(K_n-A_nJ_n)^2(s,\theta))ds \nonumber\\
\leq \int_t^{t_i}\max\limits_{S^1}e^C(e^{2\beta_n-4\mu_n}(J^2_n+2|J_n|+C^2)(s,\bar{\theta}))ds  \label{4.41}
  \end{equation}
  is bounded according to lemma $\ref{lem4.4}$.
\begin{proposition} \label{propo1} The  first derivatives of $\mu_n,~A_n,~\phi_n~$ are bounded on
$(t_0,t_1]\times S^1$.
\end{proposition}
 \textbf{Proof}: Recall expressions
 $\Upsilon_n,~\Upsilon_n^{\times},~\phi_{\xi_n},~\phi_{\varsigma_n}$. Then
 \begin{eqnarray}
  &~& (\partial_t\pm\sqrt{\alpha_{_{n-1}}}\partial_\theta)(\Upsilon_n\mp\Upsilon_n^{\times}) = -\frac{2}{t}(\dot{\mu}_{n-1}^2+\frac{e^{4\mu_{n-1}}}{4t^2}\alpha_{n-1} {A'}_{n-1}^2)+\frac{\dot{\alpha}_{n-1}}{\alpha_{n-1}}(\Upsilon_{n-1}\mp\Upsilon_{n-1}^{\times})
 \nonumber \\
   ~ &~& +2(\dot{\mu}_{n-1}\mp\sqrt{\alpha_{_{n-1}}}{\mu'}_{n-1})(\frac{e^{\beta_{n-1}-4\mu{n-1}}}{2t^2}
   J_{n-1}^2 + 2e^{2(\beta_{n-1}-\mu_{n-1})}V(\phi_{n-1}))\pm\frac{\Upsilon_{n-1}^{\times}}{t} \nonumber \\
   ~ &~& +8\pi\frac{\sqrt{\alpha_{n-1}}e^{\beta_{n-1}-4\mu_{n-1}}}{2t}\int_{\mathbb{R}^3}\frac{f_{n}(1+2e^{-2\mu_{n-1}}v_2^2)}
   {|v_0|}dv_1dv_2dv_3 \nonumber \\
   ~ &~& +\frac{e^{4\mu_{n-1}}}{4t^2}(\dot{A}_{n-1}-\sqrt{\alpha_{n-1}}{A'}_{n-1})(\frac{e^{\beta_{n-1}-4\mu_{n-1}}}{t^2}
   J_{n-1}(K_{n-1}-A_{n-1}J_{n-1})]  \nonumber \\
   ~ &~&+16\pi \frac{\sqrt{\alpha_{n-1}}e^{2\beta_{n-1}-4\mu_{n-1}}}{2t}\int_{\mathbb{R}^3}\frac{f_nv_2(v_3-A_{n-1}v_2)}
   {|v_0|}dv_1dv_2dv_3)  \label{4.42}
 \end{eqnarray}
 Let $(t_k;\theta_k)\in [t_0,t_1]\times S^1$.
 Consider the characteristic curves\\ $\gamma_n^{\pm} : \  t\mapsto t\pm \theta\sqrt{\alpha_{_{n-1}}} $ starting to $(t_k;\theta_k)$ and ending  at $t=t_1$. Then
 \begin{equation}\label{4.43}
    \int_{\gamma^-_n}(\partial_t-\sqrt{\alpha_{_{n-1}}}\partial_\theta)(\Upsilon_n+\Upsilon_n^{\times})+
    \int_{\gamma^+_n}(\partial_t+\sqrt{\alpha_{_{n-1}}}\partial_\theta)(\Upsilon_n-\Upsilon_n^{\times})=
    \int_{\gamma^-_n}L_{n-1}^++
    \int_{\gamma^+_n}L_{n-1}^-
 \end{equation}
 \begin{eqnarray}
  \text{where}~ L_{n-1}^\mp &=& -\frac{2}{t}(\dot{\mu}_{n-1}^2+\frac{e^{4\mu_{n-1}}}{4t^2}\alpha_{n-1} {A'}_{n-1}^2)+\frac{\dot{\alpha}_{n-1}}{\alpha_{n-1}}(\Upsilon_{n-1}\mp\Upsilon_{n-1}^{\times})
  \pm\frac{\Upsilon_{n-1}^{\times}}{t} \nonumber \\
   ~ &~& +2(\dot{\mu}_{n-1}\mp\sqrt{\alpha_{_{n-1}}}{\mu'}_{n-1})[\frac{e^{\beta_{n-1}-4\mu{n-1}}}{2t^2}
   J_{n-1}^2+2e^{2(\beta_{n-1}-\mu_{n-1})}V(\phi_{n-1}) \nonumber \\
   ~ &~& +8\pi\frac{\sqrt{\alpha_{n-1}}e^{\beta_{n-1}-4\mu_{n-1}}}{2t}\int_{\mathbb{R}^3}\frac{f_{n}
   (1+2e^{-2\mu_{n-1}}v_2^2)}{|v_0|}dv_1dv_2dv_3 ]\nonumber \\
   ~ &~& +\frac{e^{4\mu_{n-1}}}{4t^2}(\dot{A}_{n-1}\mp\sqrt{\alpha_{n-1}}{A'}_{n-1})[\frac{e^{\beta_{n-1}
   -4\mu_{n-1}}}{t^2}J_{n-1}(K_{n-1}-A_{n-1}J_{n-1})  \nonumber \\
   ~ &~&+16\pi \frac{\sqrt{\alpha_{n-1}}e^{2\beta_{n-1}-4\mu_{n-1}}}{2t}\int_{\mathbb{R}^3}\frac{f_nv_2
   (v_3-A_{n-1}v_2)}{|v_0|}dv_1dv_2dv_3]  \label{4.44}
 \end{eqnarray}
Since $|\Upsilon^\times_n|\leq \Upsilon_n$, $(\ref{4.43})$ becomes
\begin{eqnarray}
      &\Upsilon_n(t_k;\theta_k) = \frac{1}{2}\left[\Upsilon_n(t_i;\theta_+)
      -\Upsilon^\times_n(t_i;\theta_+)+\Upsilon_n(t_i;\theta_-)-\Upsilon^\times_n(t_i;\theta_-)\right]
      \nonumber\\
     &-\int_{\gamma^+_{n-1}}L_{n-1}^--\int_{\gamma^-_{n-1}}L_{n-1}^+ \nonumber\\
     &\leq \Upsilon_n(t_i;\theta_+)+\Upsilon_n(t_i;\theta_-)
      +\frac{1}{2}\left[\int_{\gamma^+_{n-1}}|L_{n-1}^-|+\int_{\gamma^-_{n-1}}|L_{n-1}^+| \right] \label{4.45}
\end{eqnarray}
From $(\ref{4.6})$, one have
\begin{eqnarray}
  \left|\frac{\dot{\alpha}_{_{n-1}}}{2t\alpha_{_{n-1}}}\right| &=\frac{e^{\beta_{n-1}-4\mu{n-1}}}{2t^2}J_{n-1}^2+2e^{2(\beta_{n-1}-\mu_{n-1})}V(\phi_{n-1})\nonumber\\
  &+8\pi\frac{\sqrt{\alpha_{n-1}}e^{\beta_{n-1}-4\mu_{n-1}}}{2t}\int_{\mathbb{R}^3}
  \frac{1+2e^{-2\mu_{n-1}}v_2^2}{|v_0|}f_{n}
  dv_1dv_2dv_3 \label{4.46}  \\
  \text{and} \ \
  \left|\frac{\dot{\alpha}_{_{n-1}}}{2t\alpha_{_{n-1}}}\right|
  &\geq\frac{e^{\beta_{n-1}-4\mu_{n-1}}}{t^2}J_{n-1}(K_{n-1}-A_{n-1}J_{n-1})\nonumber
  \\
  &+8\pi \frac{\sqrt{\alpha_{n-1}}e^{2\beta_{n-1}-4\mu_{n-1}}}{t}\int_{\mathbb{R}^3}\frac{v_2(v_3-A_{n-1}v_2)}{|v_0|}
  f_n dv_1dv_2dv_3 \label{4.47}
\end{eqnarray}
Otherwise
\begin{eqnarray*}
  \dot{\mu}_{n-1}\pm\sqrt{\alpha_{_{n-1}}}{\mu'}_{_{n-1}}&+&\frac{e^{2\mu_{_{n-1}}}}{2t}(\dot{A}_{n-1}\pm
  \sqrt{\alpha_{_{n-1}}}{A'}_{_{n-1}})
  \leq  (\dot{\mu}_{n-1}\pm\sqrt{\alpha_{_{n-1}}}{\mu'}_{_{n-1}})^2 \\
  &+& \frac{e^{4\mu_{_{n-1}}}}{4t^2}(\dot{A}_{n-1}\pm\sqrt{\alpha_{_{n-1}}}{A'}_{_{n-1}})^2 +\frac{1}{2} \\
  ~ &\leq& 2(\Upsilon_{n-1}+\Upsilon^{\times}_{n-1}) +\frac{1}{2}
  \leq4  \Upsilon_{n-1}+ \frac{1}{2}
\end{eqnarray*}
Therefore
  \begin{equation}\label{4.48}
   |L_{n-1}^\pm|\leq
   \left|\frac{\dot{\alpha}_{_{n-1}}}{\alpha_{_{n-1}}}\right|\left(
   2\Upsilon_{_{n-1}}+\frac{2\Upsilon_{_{n-1}}}{t}+\frac{1}{4t}\right)+\frac{3\Upsilon_{_{n-1}}}{t}
  \end{equation}
Now set
\begin{equation}\label{4.49}
       B_n(t,\theta)=16\pi
       e^{2\beta_{_{n-1}}-2\mu_{_{n-1}}}\int_{\mathbb{\mathbb{R}}^3}\frac{1+e^{-2\mu_{_{n-1}}}v_2^2
       +t^{-2}e^{2\mu_{_{n-1}}}(v_3-A_{_{n-1}}
       v_2)^2)}{|v_0|}f_n dv_1dv_2dv_3
\end{equation}
 then
\begin{eqnarray}
B_n(t,\theta) &\leq& 16\pi e^{2\beta_{_{n-1}}-2\mu_{_{n-1}}}\int_{\mathbb{R}^3}\frac{1+e^{-2\mu_{_{n-1}}}v_2^2+t^{-2}e^{2\mu_{_{n-1}}}(v_3-
A_{_{n-1}}v_2)^2}{e^{\beta_{_{n-1}}-\mu_{_{n-1}}}\sqrt{1+e^{-2\beta_{_{n-1}}+2\mu_{_{n-1}}}}u_1^2}f_ndu_1dv_2dv_3, \nonumber \\
  ~ &\leq & 16\pi e^{\beta_{_{n-1}}-\mu_{_{n-1}}}\|f_{_0}\|_{\infty}\left(1+e^{-2\mu_{_{n-1}}}\bar{v}_2+\frac{e^{2\mu_{_{n-1}}}}{t^2}
(\bar{v}_3+|A_{_{n-1}}|\bar{v}_2)\right)\bar{v}_2\bar{v}_{3}\nonumber \\
~~&\times&\int_{-\overline{u}_1}^{\overline{u}_1}\frac{du_1}{\sqrt{1+e^{-2\beta_{_{n-1}}+2\mu_{_{n-1}}}u_1^2}} \nonumber \\
 ~ &\leq& 16\pi e^{\beta_{_{n-1}}-\mu_{_{n-1}}}\|f_{_0}\|_{\infty}\left(1+e^{-2\mu_{_{n-1}}}\bar{v}_2^2+\frac{e^{2\mu_{_{n-1}}}}
 {t^2}(\bar{v}_3+|A_{_{n-1}}|\bar{v}_2)\right)\bar{v}_2\bar{v}_{3}\nonumber \\
 ~ &~&
\times2\left(e^{\beta_{_{n-1}}-\mu_{_{n-1}}}ln\left(\overline{u}_1+\sqrt{e^{2\beta_{_{n-1}}-2\mu_{_{n-1}}}
+u_1^2}\right)+e^{-1}\right)
      \label{4.50}
\end{eqnarray}
Combine lemmas $\ref{lem4.2}~and~\ref{lem4.3}$ with
$(\ref{4.45}),~(\ref{4.48})~and~(\ref{4.50})$ to obtain
\begin{eqnarray}
\sup\limits_{S^1} \Upsilon_n(t_k,\theta) &\leq&\int_{t_0}^{t_k}C\left\{\left[e^{-1}+ln\max(\overline{u}_1;e^{\beta_{_{n-1}}-\mu_{_{n-1}}})\right]
\left(2\Upsilon_{n-1}+\frac{2\Upsilon_{n-1}}{t_0}+\frac{1}{4t_0}\right)+\frac{3\Upsilon_{n-1}}{t_0}\right\}dt\nonumber\\
~&+&\int_{t_0}^{t_k}[\frac{\sup_{S^1}[e^{2\beta_{n-1}(t,\theta)-4\mu_{n-1}(t,\theta)}J^2_{n-1}(t,\theta)]}{t}
\left(2\Upsilon_{n-1}+\frac{2\Upsilon_{n-1}}{t_0}+\frac{1}{4t_0}\right)dt\nonumber \\
       ~&+&\int_{t_0}^{t_k}\frac{\sup_{S^1}[e^{2\beta_{n-1}(t,\theta)}(K_{n-1}-A_{n-1}J_{n-1})^2(t,\theta)]}
{t^3}\left(2\Upsilon_{n-1}+\frac{2\Upsilon_{n-1}}{t_0}+\frac{1}{4t_0}\right)dt \nonumber \\
       ~&+&  \sup\limits_{S^1} \Upsilon_n(t_0,\theta) \label{4.50+}
     \end{eqnarray}
Otherwise,
 \begin{eqnarray}
       (u_1(t_{_k},\theta_{_k}))^2 &\leq& (\bar{u}_1(t_{_0}))^2+C\int_{t_{_0}}^{t_{_k}}\{\sup_{S^1}[e^{2\beta_{n-1}-4\mu_{n-1}}(J_{n-1}(t,\theta))^2] \nonumber\\
       ~~&+&\sup_{S^1}\frac{[e^{2\beta_{n-1}}((K_{n-1}-A_{n-1}J_{n-1})(t,\theta))^2]}{t^3} \nonumber\\
       ~~ &+&(e^{-1}+e^{\beta_{n-1}-\mu_{n-1}}ln(\sup_{S^1}\{\bar{u}_1;e^{\beta_{n-1}-\mu_{n-1}}\})(\bar{u}_1(t))^2\nonumber\\
       ~~&+&(\Upsilon_{n-1}+\bar{u}^2_1+1)\{\sup_{S^1} e^{2\beta_{n-1}-2\mu_{n-1}}+(\sup e^{2\beta_{n-1}-2\mu_{n-1}})\bar{v}^2_2\nonumber\\
       ~~ &+& \frac{[\sup_{S^1} e^{\beta_{n-1}}\bar{v}_3+\sup_{S^1}
       (e^{\beta_{n-1}}|A_{n-1}|\bar{v}_2)]^2}{t^2_0}\nonumber\\
       ~~&+&\frac{[\sup_{S^1} e^{\beta_{n-1}-\mu_{n-1}}\bar{v}_3+\sup_{S^1}(e^{\beta_{n-1}-\mu_{n-1}}|A|\bar{v}_2)]\bar{v}_2)}{t_0}\nonumber\\
       ~~ &+& \frac{[\sup_{S^1} e^{\beta_{n-1}}\bar{v}_3+\sup_{S^1}(e^{\beta_{n-1}}|A|\bar{v}_2)]\max_{[0,1]}(e^{\beta_{n-1}}(K_{n-1}-A_{n-1}J_{n-1}))}{t_f}\nonumber\\
       ~~ &+&\frac{\sup_{S^1} (e^{\beta_{n-1}-2\mu_{n-1}})\sup_{S^1}(e^{\beta_{n-1}-2\mu_{n-1}}|J_{n-1}|)\bar{v}_2}{t}
       \}dt \label{4.51}
 \end{eqnarray}
Using  $(\ref{4.46})-(\ref{4.47})$  and lemmas
$\ref{lem4.4}~and~\ref{lem4.5}$, we deduce that the right hand side of  $(\ref{4.51})$ is bounded.\\
$~~$ To conclude the proof of this proposition, we need two others lemmas:
\begin{lemma}\label{lem4.6}
For all $t\in (t_0,t_1]~and~ \theta \in S^1$, we have
\begin{eqnarray}
  \chi_n(t) &\leq& C+C(t)\int_{t_0}^t [\chi_{n-1}(s)(1+\bar{u}_{n-1,1}ln\bar{u}_1+ln\bar{u}_{n-1,1}(s))\nonumber\\
  &+&\underset{\theta\in S^1}{sup}\Upsilon_{n-1}(s)(\chi_{n-1}(s))^2]ds
  +2C'\underset{\bar{t}\in\{t_0,t\}}{max}|V(\phi_{n-1}(\bar{t},\theta))|\label{4.52}
\end{eqnarray}
where $ \chi_n(t)$ is defined like $(\ref{85'+'})$ in which   $\xi~and~\varsigma$ are respectively substituted by  $\xi_n~and~\varsigma_n$.
\end{lemma}
\textbf{Proof}:
 Integrating $(\ref{4.8})~and~ (\ref{4.9})$  along null paths $\partial_{\xi_{n-1}}~and~\partial_{\varsigma_{n-1}}$
from $t=t_0$ to $(t,\theta)\in [t_0,t_1)\times S^1$ gives respectively
\begin{eqnarray}
  \phi_{\varsigma_n}(t,\theta) &=& \phi_{\varsigma_n}(t_0,\theta-(N_{n-1}(t)-t_0))+\int_{t_0}^t\frac{\dot{\alpha}_{_{n-1}}}{2\alpha_{_{n-1}}}
  \phi_{\varsigma_{n-1}}(s,\theta-(N_{n-1}(s)-t_0))ds \nonumber \\
  ~~&+&\int_{t_0}^t\left[\alpha_{_{n-1}}(2A'_{_{n-1}} H_{_{n-1}}(G_{_{n-1}}-1)-2\mu'_{_{n-1}}(G_{_{n-1}}+A_{_{n-1}}H_{_{n-1}})^2e^{-2(\tau_{_{n-1}}
  -2\mu_{_{n-1}})})\right]\nonumber \\
  ~~&\times&\phi'_{_{n-1}}(s,\theta-(N_{_{n-1}}(s)-t_0))ds -\int_{t_0}^t(4\alpha_{_{n-1}}\mu'_{_{n-1}}
  H_{_{n-1}}^2s^2e^{-2\tau_{_{n-1}}}\phi'_{_{n-1}}\nonumber \\
  ~~&+&\frac{\dot{\phi}_{_{n-1}}}{s}+e^{2\beta_{_{n-1}}-2\mu_{_{n-1}}} V'(\phi_{_{n-1}}))(s,\theta-(N_{_{n-1}}(s)-t_0))ds \label{4.53}
  \end{eqnarray}
  \begin{eqnarray}
   \phi_{\xi_n}(t,\theta) &=& \phi_{\xi_n}(t_0,\theta+(N_{_{n-1}}(t)-t_0))+\int_{t_0}^t\frac{\dot{\alpha}_{_{n-1}}}{2\alpha_{_{n-1}}}
   \phi_{\xi_{_{n-1}}}(s,\theta+(N_{_{n-1}}(s)-t_0))ds \nonumber \\
   ~~&+&\int_{t_0}^t\left[\alpha_{_{n-1}}(2A'_{_{n-1}} H_{_{n-1}}(G_{_{n-1}}-1)-2\mu'_{_{n-1}}(G_{_{n-1}}+A_{_{n-1}}H_{_{n-1}})^2e^{-2(\tau_{_{n-1}}
   -2\mu_{_{n-1}})}\right]\nonumber \\
   ~&\times&\phi'_{_{n-1}}(s,\theta+(N_{_{n-1}}(s)-t_0))ds -\int_{t_0}^t(4\alpha_{_{n-1}}\mu'_{_{n-1}}
  H_{_{n-1}}^2s^2e^{-2\tau_{_{n-1}}}\phi'_{_{n-1}}\nonumber\\
  ~&+&\frac{\dot{\phi}_{_{n-1}}}{s}+e^{2\beta_{_{n-1}}-2\mu_{_{n-1}}} V'(\phi_{_{n-1}}))(s,\theta+(N_{_{n-1}}(s)-t_0))ds \label{4.54}
\end{eqnarray}
Analogously as in $(\ref{4.50})$ and using lemmas
$\ref{lem4.4}-\ref{lem4.5}$, $\left|\frac{\dot{\alpha}_{_{n-1}}}{\alpha_{_{n-1}}}\right|~$
is bounded by $ln(1+\bar{u}_{n-1,1}^2)$ and we obtain respectively
\begin{eqnarray}
  \underset{\theta \in S^1}{sup}|\phi_{\varsigma_n}|(t,\theta) &\leq \underset{\theta \in S^1}{sup}|\phi_{\varsigma_n}(t_0,\theta)|+\int_{t_0}^tC(s)[\underset{\theta \in S^1}{sup}|\phi_{\varsigma_{n-1}}|(s,\theta)ln(1+\overline{u}_{n-1,1}^2(s))\nonumber \\
&+\frac{1}{s}\underset{\theta\in S^1}{sup}|\dot{\phi}_{n-1}(s,\theta)|+e^{2\beta_{n-1}-2\mu_{n-1}}|V'(\phi_{n-1})|]ds+\kappa_{n-1}(t,\theta) \nonumber \\
&\leq C+ \kappa_{n-1}(t,\theta)+ \int_{t_0}^t C(s)[\underset{\theta \in S^1}{sup}|\phi_{\varsigma_{n-1}}|(s,\theta)ln(1+\overline{u}_{_{n-1,1}}^2(s))
 \nonumber \\
&+ \frac{1}{s}\underset{\theta\in S^1}{sup}|\dot{\phi}_{_{n-1}}(s,\theta)]ds+\widetilde{C} \left|\int_{t_0}^t\dot{\phi}_{_{n-1}}(s,\theta)V'(\phi_{n-1}(s,\theta)ds\right| \nonumber \\
&\leq C+ \kappa_{n-1}(t,\theta)+\int_{t_0}^tC(s)[\underset{\theta \in S^1}{sup}|\phi_{\varsigma_{n-1}}|(s,\theta)ln(1+\overline{u}_{_{n-1,1}}^2(s)) + \nonumber \\
&+ \frac{1}{s}\underset{\theta\in S^1}{sup}|\dot{\phi}_{_{n-1}}(s,\theta)|]ds+\widetilde{C} |V(\phi_{n-1}(t,\theta))-V(\phi_{n-1}(t_0,\theta))| \label{4.55}
  \end{eqnarray}
  \begin{eqnarray}
 \underset{\theta \in S^1}{sup}|\phi_{\xi_n}|(t,\theta) &\leq \underset{\theta \in S^1}{sup}|\phi_{\xi_n}(t_0,\theta)|+\int_{t_0}^tC(s)[\underset{\theta \in S^1}{sup}|\phi_{\xi_{n-1}}|(s,\theta)ln(1+\overline{u}_{_{n-1,1}}^2(s)) \nonumber \\
 &+ \frac{1}{s}\underset{\theta\in S^1}{sup}|\dot{\phi}_{n-1}(s,\theta)|+e^{2\beta_{n-1}-2\mu_{n-1}}|V'(\phi_{n-1})|]ds +\tilde{\kappa}_{n-1}(t,\theta)\nonumber \\
 &\leq C +\tilde{\kappa}_{n-1}(t,\theta)+ \int_{t_0}^tC(s)[\underset{\theta \in S^1}{sup}|\phi_{\xi_{n-1}}|(s,\theta)ln(1+\overline{u}_{_{n-1,1}}^2(s)) \nonumber \\
 &+ \frac{1}{s}\underset{\theta\in S^1}{sup}|\dot{\phi}_{_{n-1}}(s,\theta)]ds+\widetilde{C} \left|\int_{t_0}^t\dot{\phi}_{_{n-1}}(s,\theta)V'(\phi_{n-1}(s,\theta)ds\right| \nonumber \\
 \leq& C +\tilde{\kappa}_{n-1}(t,\theta) + \int_{t_0}^tC(s)[\underset{\theta \in S^1}{sup}|\phi_{\xi_{n-1}}|(s,\theta)ln(1+\overline{u}_{_{n-1,1}}^2(s)) \nonumber \\
 +&\frac{1}{s}\underset{\theta\in S^1}{sup}|\dot{\phi}_{_{n-1}}(s,\theta)|]ds+\widetilde{C} |V(\phi_{n-1}(t,\theta))-V(\phi_{n-1}(t_0,\theta))|  \label{4.56}
\end{eqnarray}
where $\kappa_{n-1}~and~\tilde{\kappa}_{n-1}$ are defined like
\begin{eqnarray*}
  \kappa(t,\theta) &=& \int_{t_0}^t\alpha\left[(2A_\theta
H(G-1)-2\mu_\theta(G+AH)^2)e^{-2(\tau-2\mu)}-4\mu_\theta
  H^2s^2e^{-2\tau}\right]\nonumber \\
  ~~&~&.\phi_\theta(s,\theta-(N(s)-t_0))ds \label{kappa1}\\
  \tilde{\kappa}(t,\theta) &=& \int_{t_0}^t\alpha\left[(2A_\theta
H(G-1)-2\mu_\theta(G+AH)^2)e^{-2(\tau-2\mu)}-4\mu_\theta
  H^2s^2e^{-2\tau}\right]\nonumber \\
  ~~&~&. \phi_\theta(s,\theta+(N(s)-t_0))ds\label{kappa2}
\end{eqnarray*}
in which 
\begin{equation*}
A,~\tau,~\mu,~\alpha,~G,~H,~A_\theta,~\mu_\theta,~\phi_\theta,~\phi_t
\end{equation*}
are respectively replaced by
\begin{equation*}
    A_{n-1},~\tau_{n-1},~\mu_{n-1},~\alpha_{n-1},~G_{n-1},~H_{n-1},~A'_{n-1},~\mu'_{n-1},~\phi'_{n-1},~\dot{\phi}_{n-1}
\end{equation*}
$\kappa_{n-1}(t,\theta)~and~\tilde{\kappa}_{n-1}(t,\theta)$ are bounded in  similar way  of
$\kappa(t,\theta)~and~\tilde{\kappa}(t,\theta)$ using $(\ref{En})$. 
Estimation of $\beta_{n-1}$ allows us to bound
\begin{equation*}
\phi'_{n-1};~A'_{n-1},~e^{-2(\tau_{n-1}-2\mu_{n-1})}~and~\mu'_{n-1}.
\end{equation*}
 \begin{equation*}
\text{Since}~(K_{n-1}-A_{n-1}J_{n-1})^2\lesssim \mathcal{E}^{^\circ}
~and~K_{n-1}-A_{n-1}J_{n-1}=-\frac{t^3e^{-2\tau_{n-1}}}{\sqrt{\alpha_{n-1}}}\dot{H}_{_{n-1}},
\end{equation*}  we deduce that
 $\dot{H}_{_{n-1}}$ and $H_{n-1}$ are bounded.
%
Now from 
 \begin{eqnarray}
 \sqrt{\alpha_{_{n-1,2}}}S_{n-1,1k} &=& \int_{\mathbb{R}^3}f_n\frac{v^1v^k}{v^0}dv ~~k\in{2;3} \nonumber\\
  ~~ &\lesssim&\|f_0\|_\infty
  \int_{\mathbb{R}}\sup{|v^1|}\frac{dv^1}{\sqrt{1+(v^1)^2}} \nonumber\\
  ~~&\leq& C\bar{u}_{_{n-1,1}}ln(\bar{u}_{_{n-1,1}})\label{4.56'},
\end{eqnarray}
 $\Gamma_{n-1}=\dot{G}_{_{n-1}}+A_{n-1}\dot{H}_{n-1}$ are bounded and then  $\dot{G}_{_{n-1}}$
and $G_{_{n-1}}$ are bounded.
We deduce from expressions of $\kappa_{n-1}$ and $\tilde{\kappa}_{n-1}$, that   
\begin{eqnarray}
|\kappa_{n-1}(t,\theta)| &\leq& C\int_{t_0}^t C\bar{u}_{_{n-1,1}}ln(\bar{u}_{_{n-1,1}})(s,\theta -(N_{n-1}(s)-t_0))ds \label{4.57}\\
|\tilde{\kappa}_{n-1}(t,\theta)|&\leq&
  C\int_{t_0}^tC\bar{u}_{_{n-1,1}}ln(\bar{u}_{_{n-1,1}})(s,\theta +(N_{n-1}(s)-t_0))ds \label{4.58}
\end{eqnarray}
Add $(\ref{4.55}),~(\ref{4.56}),~(\ref{4.57})~and~(\ref{4.58})$ to
obtain  $(\ref{4.52})$.
\begin{lemma}\label{lem4.7}
For all $t\in (t_0,t_i]~,~ \theta \in S^1$, we have
\begin{equation}\label{4.59}
    \Lambda_{_{n}}(t;\theta)\leq
    C+C(t)\int_{[t_0,t]}(\Lambda_{_{n-1}}(s;\theta))^3
    ln(\Lambda_{_{n-1}}(s;\theta))ds
\end{equation}
\end{lemma}
\textbf{Proof}:
Inequalities $(\ref{4.50})~and~(\ref{4.52})$ give
\begin{equation} \label{4.60}
\psi_n(t) \leq
 C+\int_{t_0}^t\varrho_{n-1}(s)\psi_{n-1}(s)ln(\psi_{n-1}(s))ds
\end{equation}
where $C$ is a nonnegative constant depending on the initial data at time $t_0$ and
\begin{eqnarray*}
&\varrho_{n-1}(s)= 2se^{2\beta_{n-1}-\mu_{n-1}}2V(\phi_{n-1})\\
+&2se^{2\beta_{n-1}-\mu_{n-1}}16\pi\left(\frac{\sqrt{\alpha_{n-1}}}{t}\int_{\mathbb{R}^3}f_n\frac{1+e^{-2\mu_{n-1}}
v_2^2+e^{2\mu_{n-1}}t^2(v_3-A_{n-1}v_2)^2}{|v_0|}dv\right)
\end{eqnarray*}
Adding  $(\ref{4.52}),(\ref{4.60})$ and using the fact that
\begin{equation*}
    \chi_{n-1}(s)+\psi_{n-1}(s)\chi^2_{n-1}(s)\leq \Lambda_{n-1}(s)+\Lambda^3_{n-1}(s)
    \lesssim \Lambda^3_{n-1}(s)
\end{equation*}
give $(\ref{4.59})$.
\begin{lemma}\label{lem4.8}
For all $(\theta;t)\in S^1\times (t_0,t_1)$,
$\partial_t f_n(t,\theta,.),~\partial_\theta f_n(t,\theta,.)$ are bounded.
\end{lemma}
\textbf{Proof}: For $z=t~or~\theta$ one have:
\begin{equation}  \label{4.61'}
    \frac{\partial f_n}{\partial z}=\frac{\partial f_0}{\partial \theta} \frac{\partial \Theta_n}{\partial
    z}+\frac{\partial f_0}{\partial v} \frac{\partial V_n}{\partial z}
\end{equation}
with\\
$\left\{%
\begin{array}{llll}
\strut
 \frac{d \Theta_n}{ds} = \frac{V^1_n\sqrt{\alpha_{n-1}}}{V^0}~~~~~~~~~~~~~~~~~~~~~~~~~~~~~~~~~~~~~~~~~~~~~~~~~~~~~~~~~~~~~~~~~~~~~~~~~~~~~~~\label{4.62} \\
 \frac{d V^1_n}{ds} =
 -(\tau'_{n-1}-\mu'_{n-1}+\frac{\alpha'_{n-1}}{2\alpha_{n-1}})\sqrt{\alpha_{n-1}}V^0-(\dot{\tau}_{n-1}
 -\dot{\mu}_{n-1})V^1_n \nonumber \\
     ~~~~~-\frac{\sqrt{\alpha_{n-1}}\mu'_{n-1}}{V^0}((V^2_n)^2-(V^3_n)^2)
     +\frac{\sqrt{\alpha_{n-1}}A'_{n-1}}{sV^0}e^{2\mu_{n-1}}V^2_nV^3_n \\
\frac{dV^2_n}{ds}=-\dot{\mu}_{n-1}V^2_n-\sqrt{\alpha_{n-1}}\mu'_{n-1}\frac{V^1_nV^2_n}{V^0}~~~~~~~~~~~~~~~~~~~\\
\frac{d V^3_n}{ds} =-\left(\frac{1}{s}-\dot{\mu}_{n-1}\right)V^3_n+\sqrt{\alpha_{n-1}}\mu'_{n-1}\frac{
 V^1_nV^3_n}{V^0}-\frac{e^{2\mu_{n-1}}}{s}\left(\dot{A}_{n-1}+\sqrt{\alpha_{n-1}}A'_{n-1}\frac{V^1_n}{V^0}\right)V^2_n~~~~~~~~~~~\\
\end{array}%
\right.$\\
 Let \
 $\Theta_n(s)=\Theta_n(s,t,\theta,v)~and~V^k_n(s)=V^k_n(s,t,\theta,v),~k=1,2,3$ be a solution of the previous characteristic system 
  and $\partial ~~\text{being either}~~ \partial_t,~\partial_\theta,~or~\partial_{v^k}$. Define
  \begin{eqnarray}
   \Psi_n &=& \alpha^{-1/2}_{n-1}\partial\Theta_n, \label{4.66} \\
   Z^1_n &=& \partial V^1_n+\left(\frac{\dot{\tau}_{n-1}V^0}{\sqrt{\alpha_{n-1}}}-\frac{\dot{\mu}_{_{n-1}}V^0}{\sqrt{\alpha_{n-1}}}
   \frac{(V^0)^2-(V^1_n)^2+(V^2_n)^2-(V^3_n)2}{(V^0)^2-(V^1_n)^2}\right)\partial\Theta_n\nonumber  \\
   ~ &~&+\left(\mu'_{_{n-1}}\frac{V^1_n((V^2_n)^2-(V^3_n)^2)}{(V^0)^2-(V^1_n)^2}-\frac{\dot{A}_{_{n-1}}
   e^{2\mu_{n-1}}}{\sqrt{\alpha_{n-1}}t}\frac{V^0V^2_nV^3_n}{(V^0)^2-(V^1_n)^2}\right)\partial\Theta_n  \nonumber\\
   ~ &~&+\left(A_\theta \frac{V^1_nV^2_nV^3_n}{(V^0)^2-(V^1_n)^2}\right)\partial\Theta_n, \label{4.67}\\
   Z^2_n &=&\partial V^2_{_{n}}+V^2_{_{n}}\mu'_{_{n-1}} \partial\Theta_n, \label{4.68} \\
   Z^3_n &=& \partial V^3_{_{n}}-(V^3_{_{n}}\mu'_{_{n-1}}-\frac{e^{2\mu_{_{n-1}}}}{s}A'_{_{n-1}}) \partial\Theta_n\label{4.69}
 \end{eqnarray}
  Then there are two matrix  $M_n=\{a_{\varsigma\upsilon}^n\},~\left(\varsigma,\upsilon\in \{0,1,2,3\}\right)$ and
  $\Omega_n:=(\Psi_n,Z^1_n,Z^2_n,Z^3_n)^T$
satisfying
\begin{equation}\label{4.70}
    \frac{d\Omega_n}{ds}=M_{_{n-1}}\Omega_n,
\end{equation}
with
$a_{\varsigma\upsilon}^n=a_{\varsigma\upsilon}(s,\Theta_n(s),V^k_n(s))$  all uniformly bounded on $[t_0,t_1)$.
Particularly
\begin{eqnarray}
  \frac{dZ^2_n}{ds} &=& a_{22}^{n-1}Z^2_n \label{4.71}\\
  \frac{dZ^3_n}{ds} &=&a_{30}^{n-1}\Psi_n +a_{32}^{n-1}Z^2_n+a_{33}^{n-1}Z^3_n \label{4.72}
\end{eqnarray}
 $a_{_{30}}^{n-1}$, $a_{_{33}}^{n-1}~$ and $a_{32}^{n-1}$ depend respectively on
$\frac{\dot{\alpha}_{n-1}}{\alpha_{n-1}}$, $-\frac{1}{s}+\dot{\mu}_{n-1}$ and \\
$\frac{e^{2\mu_{n-1}}}{s}(\dot{A}_{n-1}+\sqrt{\alpha_{n-1}}A'_{n-1}\frac{V^1_n}{V^0})$, which are uniformly bounded on $[t_0, t_1)$.
Equation $(\ref{4.67})$ comes from $(\ref{4.12})-(\ref{4.13}), \ (\ref{4.5})$ and $(\ref{4.7})$.
Otherwise
\begin{eqnarray*}
  \left|\frac{d}{ds}\partial_\theta(\Theta,V)_{n+1}(s,t,\theta,v)\right| &=& \left|\partial_\theta(\Theta,V)_{n+1}(s,t,\theta,v)\partial_\theta\Xi_n(s,\Theta(s,t,\theta,v),V(s,t,\theta,v))\right| \\
   ~ &\leq&
  |(\Theta,V)_{n+1}(s,t,\theta,v^1,v^2,v^3)|(C(s)+\Lambda_n(s))|
\end{eqnarray*}
therefore, for $(s,\theta,v^1,v^2,v^3)\in
suppf_{n+1}(t)\cup suppf_{n}(t)$,
 Gronwall's inequality gives
\begin{equation*}
    |\partial_\theta(\Theta,V)_{n+1}(t_0,t,\theta,v^1,v^2,v^3)|\leq
    exp\left\{\int_{t_0}^t|(C(s)+\Lambda_n(s))|ds\right\}
\end{equation*}
Relation $(\ref{4.61'})$  allows us to bounded the derivative of
 $f_n$ since all components of  $a_{\varsigma\upsilon}^{n-1}$
are bounded. One deduce (see $(\ref{4.17n})$) that\\
$\|\partial_\theta f_{n+1}(t)\|=
\|\partial_{(\theta,v)}(\Theta,V)_{n+1}(t_0,t,\theta,v^1,v^2,v^3)f_0((\Theta,V)_{n+1}(t_0,t,\theta,v^1,v^2,v^3))\|$
and
\begin{equation}\label{4.73.}
    \|\partial_\theta f_{n+1}(t)\|\leq  \|\partial_{(\theta,v)} f_0\|
    exp\left(\int_{t_0}^t C(s)(D_n(s)+\Lambda_n(s))ds\right)
\end{equation}
We also deduce from  lemmas $\ref{lem4.6} ~\text{and}~\ref{lem4.7}$, bounds of
$\mu_n,~\alpha_n,~A_n,~\tau_n,~\phi_n,~G_n,~H_n,~f_n$ and 
of
\begin{equation*}
\dot{\mu}_n,~\dot{\alpha}_n,~\dot{A}_n,~\dot{\tau}_n,~\dot{\phi}_n,~\dot{H}_n,~\mu'_n,~\alpha'_n,~A'_n,
~\tau'_n,~\phi'_n,~\dot{G}_n,~f'_n,~G'_n,~H'_n.
\end{equation*}
This completes the proof of the proposition.
\begin{proposition}\label{propo3} Let $[t_0;T^{*}]\subset [t_0;T]$ be an arbitrary compact subset on which the
previous estimates hold. Then on such an interval, the iterates converge uniformly
for $L^\infty-norm$.
\end{proposition}
\textbf{Proof}: Let $t\in [t_0;T^{*}]$ and
 \begin{eqnarray}
   \digamma_n(t)&:=&\sup\{\|(f_{n+1}-f_n)(s)\|+\||(\phi_{n+1}-\phi_n)(s)\||+\||(\alpha_{n+1}-\alpha_n)(s)|\|\nonumber \\
   ~ &+& \||(\mu_{n+1}-\mu_n)(s)\||+\||(A_{n+1}-A_n)(s)\||+\||(\tau_{n+1}-\tau_n)(s)\|| \nonumber  \\
   ~ &+&
   \||(G_{n+1}-G_n)(s)|\|+\||(H_{n+1}-H_n)(s)|\|;t_0\leq s
   \leq T^*\}  \label{4.73}
 \end{eqnarray}
 where for all function $h:=h(.,\theta)$,
 \begin{equation*}
 \||(h_{n+1}-h_n)(t)\||=\|(h_{n+1}-h_n)(t)\|+\|(\dot{h}_{n+1}-\dot{h}_n)(t)\|+\|(h'_{n+1}-h'_n)(t)\|
\end{equation*}
We use this notation in what follows :
\begin{equation*}
h^{\{\pm\}}\equiv \partial_n^{\pm}h=(\partial_t\pm\sqrt{\alpha_n}\partial_\theta)h.
\end{equation*}
Proceeding analogously as in the proof of proposition
$\ref{propo1}$, and using $(\ref{4.8})-(\ref{4.9})$, we have
 \begin{eqnarray}
  &~& \partial_n^{\pm}(\bar{\Upsilon}_{n+1}\mp\bar{\Upsilon}_{n+1}^{\times})-\partial_{n-1}^{\pm}
  (\bar{\Upsilon}_{n}\mp\bar{\Upsilon}_{n}^{\times}) = +2\mu^{\{\pm\}}_{n}(T^1_n-T^1_{n-1})+ 2(\mu^{\{\pm\}}_{n}-\mu^{\{\pm\}}_{n-1})T^1_{n-1}\nonumber \\
   &~&-\frac{\dot{\alpha}_{n}}{\alpha_{n}}[\bar{\Upsilon}_{n}-\bar{\Upsilon}_{n-1}\mp(\bar{\Upsilon}_{n}^{\times}
   -\bar{\Upsilon}^\times_{n-1})]+(\frac{\dot{\alpha}_{n}}{\alpha_{n}}-\frac{\dot{\alpha}_{n-1}}{\alpha_{n-1}})
   (\bar{\Upsilon}_{n-1}\mp\bar{\Upsilon}_{n-1}^{\times}) \nonumber \\
&~& -\frac{2}{t}(\dot{\phi}_{n}^2-\dot{\phi}_{n-1}^2)+2\phi^{\{\pm\}}_{n}(\daleth_n-\daleth_{n-1})+2(\phi^{\{\pm\}}_{n}
-\phi^{\{\pm\}}_{n-1})\daleth_{n-1}\pm\left(\frac{\bar{\Upsilon}_{n}^{\times}}{t}
   -\frac{\bar{\Upsilon}_{n-1}^{\times}}{t}\right) \nonumber\\
&~&-\frac{2}{t}(\dot{\mu}_{n}^2-\dot{\mu}_{n-1}^2+\frac{e^{4\mu_{n}}}{4t^2}\alpha_{n} {A'}_{n}^2-\frac{e^{4\mu_{n-1}}}{4t^2}\alpha_{n-1} {A'}_{n-1}^2) \nonumber \\
 &~& +\frac{e^{4\mu_{n}}}{4t^2}A^{\{\pm\}}_{n}(T^2_n-T^2_{n-1})+\frac{T^2_{n-1}}{4t^2}(e^{4\mu_{n}}-e^{4\mu_{n-1}}) \label{4.75}
 \end{eqnarray}
 where
  \begin{eqnarray}
 T^1_n&=&\frac{e^{2\beta_{n}-4\mu_{n}}}{2t^2}J_{n}^2+2e^{2(\beta_{n}-\mu_{n})}V(\phi_{n})\nonumber \\
 &+&8\pi\frac{\sqrt{\alpha_{n}}e^{2\beta_{n}-4\mu_{n}}}{2t}\int_{\mathbb{R}^3}\frac{f_{n+1}(1+2e^{-2\mu_{n}}v_2^2)}{|v_0|}
 dv_1dv_2dv_3\label{4.76}\\
T^2_n&=&\frac{e^{2\beta_{n}-4\mu_{n}}}{t^2}J_{n}(K_{n}-A_{n}J_{n})\nonumber\\
&+&16\pi\frac{\sqrt{\alpha_{n}}e^{2\beta_{n}-4\mu_{n}}}{2t}\int_{\mathbb{R}^3}\frac{f_{n+1}v_2(v_3-A_{n}v_2)}{|v_0|}
dv_1dv_2dv_3)\label{4.77}
\end{eqnarray}
Using $(\ref{4.6})$  in $(\ref{4.76})-(\ref{4.77})$ gives
\begin{equation}\label{4.79}
|T^1_n-T^1_{n-1}|\leq\frac{1}{2t}\left|\frac{\dot{\alpha}_{n}}{\alpha_{n}}-\frac{\dot{\alpha}_{n-1}}
{\alpha_{n-1}}\right| \ \text{and} \  |T^2_n-T^2_{n-1}|\leq\frac{1}{2}\left|\frac{\dot{\alpha}_{n}}{\alpha_{n}}
-\frac{\dot{\alpha}_{n-1}}
{\alpha_{n-1}}\right|
\end{equation}
Next,
\begin{eqnarray}
  \int_{\gamma_n^{-}\cup \gamma_{n+1}^{-}}\left[\partial_n^{-}(\bar{\Upsilon}_{n+1}+\bar{\Upsilon}_{n+1}^{\times})-\partial_{n-1}^{-}
  (\bar{\Upsilon}_{n}+\bar{\Upsilon}_{n}^{\times})\right] &+&~~~~~~~~~~~~~~~~~~~~~~~~~~~~ \nonumber \\
  \int_{\gamma_n^{+}\cup \gamma_{n+1}^{+}}\left[\partial_n^{+}(\bar{\Upsilon}_{n+1}-\bar{\Upsilon}_{n+1}^{\times})-\partial_{n-1}^{+}
  (\bar{\Upsilon}_{n}-\bar{\Upsilon}_{n}^{\times})\right]&=&~~~~~~~~~~~~~~~~~~~~~~~~~~~~ \nonumber \\
  \int_{\gamma_n^{-}\cup
\gamma_{n+1}^{-}}\tilde{L}_n^++\int_{\gamma_n^{+}\cup
  \gamma_{n+1}^{+}}\tilde{L}_n^- &~&~~~~~~~~~~~~~~~~~~~~~~~~~~~\label{4.80}
\end{eqnarray}
where $ \tilde{L}_n^\pm$ is define using the right hand side of $(\ref{4.75})$.
\begin{eqnarray}
  \int_{\gamma_{n+1}^{-}}\left[\partial_n^{-}(\bar{\Upsilon}_{n+1}-\bar{\Upsilon}_{n}+\bar{\Upsilon}_{n+1}^{\times}
  -\bar{\Upsilon}_{n}^{\times})\right] &+&\int_{\gamma_n^{-}\cup
 \gamma_{n+1}^{-}}(\partial_{n}^--\partial_{n-1}^{-})(\bar{\Upsilon}_{n}+\bar{\Upsilon}_{n}^{\times})+ \nonumber\\
  \int_{ \gamma_{n+1}^{+}}\left[\partial_n^{+}(\bar{\Upsilon}_{n+1}-\bar{\Upsilon}_{n}+\bar{\Upsilon}_{n+1}^{\times}
  -\bar{\Upsilon}_{n}^{\times})\right] &+&\int_{\gamma_n^{+}\cup
  \gamma_{n+1}^{+}}(\partial_{n}^+-\partial_{n-1}^{+})(\bar{\Upsilon}_{n}+\bar{\Upsilon}_{n}^{\times}) \nonumber \\
  ~~&=& \int_{\gamma_n^{-}\cup
\gamma_{n+1}^{-}}\tilde{L}_n^++\int_{\gamma_n^{+}\cup
  \gamma_{n+1}^{+}}\tilde{L}_n^- \label{4.81}
\end{eqnarray}
\begin{eqnarray}
  \int_{ \gamma_{n+1}^{-}}\partial_n^{-}(\bar{\Upsilon}_{n+1}-\bar{\Upsilon}_{n})+\int_{ \gamma_{n+1}^{+}}\partial_n^{+}(\bar{\Upsilon}_{n+1}-\bar{\Upsilon}_{n}) &=&-\int_{\gamma_n^{-}\cup
  \gamma_{n+1}^{-}}(\partial_{n}^--\partial_{n-1}^{-})(\bar{\Upsilon}_{n}+\bar{\Upsilon}_{n}^{\times}) \nonumber \\
 - \int_{ \gamma_{n+1}^{+}}\left[\partial_n^{+}(\bar{\Upsilon}_{n+1}^{\times}-\bar{\Upsilon}_{n}^{\times})\right]
 &-&\int_{\gamma_n^{+}\cup
  \gamma_{n+1}^{+}}(\partial_{n}^+-\partial_{n-1}^{+})(\bar{\Upsilon}_{n}+\bar{\Upsilon}_{n}^{\times}) \nonumber \\
  ~~&+& \int_{\gamma_n^{-}\cup
\gamma_{n+1}^{-}}\tilde{L}_n^++\int_{\gamma_n^{+}\cup
  \gamma_{n+1}^{+}}\tilde{L}_n^- \label{4.82}
\end{eqnarray}
\begin{eqnarray}
    (\bar{\Upsilon}_{n+1}-\bar{\Upsilon}_{n})(t_k,\theta_k) &=& \frac{1}{2}[\bar{\Upsilon}_{n+1}-\bar{\Upsilon}_{n})(t_0,\theta_+)-(\bar{\Upsilon}^\times_{n+1}
    -\bar{\Upsilon}^\times_{n})(t_0,\theta_+)]- \nonumber\\
   \frac{1}{2}\left[ \int_{\gamma_n^{-}\cup
  \gamma_{n+1}^{-}}(\partial_{n}^--\partial_{n-1}^{-})(\bar{\Upsilon}_{n}+\bar{\Upsilon}_{n}^{\times})\right]
  &+&\frac{1}{2}[\bar{\Upsilon}_{n+1}-\bar{\Upsilon}_{n})(t_0,\theta_-)-(\bar{\Upsilon}^\times_{n+1}
  -\bar{\Upsilon}^\times_{n})(t_0,\theta_-)) ] -\nonumber \\
  \frac{1}{2}\left[\int_{\gamma_n^{+}\cup
  \gamma_{n+1}^{+}}(\partial_{n}^+-\partial_{n-1}^{+})(\bar{\Upsilon}_{n}+\bar{\Upsilon}_{n}^{\times})\right] &+&\frac{1}{2}\left[\int_{\gamma_n^{-}\cup
\gamma_{n+1}^{-}}\tilde{L}_n^++\int_{\gamma_n^{+}\cup
  \gamma_{n+1}^{+}}\tilde{L}_n^- \right] \label{4.83}
  \end{eqnarray}
  However
  \begin{eqnarray}
   &~& |\tilde{L}_n^\pm| \leq \left(\frac{2}{t}+\frac{|\dot{\alpha}_n|}{|\alpha_n|}\right)(\bar{\Upsilon}_{n}
    -\bar{\Upsilon}_{n-1})+\left|\frac{|\dot{\alpha}_n|}{|\alpha_n|}-\frac{|\dot{\alpha}_{n-1}|}
    {|\alpha_{n-1}|}\right|(\bar{\Upsilon}^\times_{n}\pm\bar{\Upsilon}^\times_{n-1}) +\nonumber \\
    &~&\left|\frac{|\dot{\alpha}_n|}{|\alpha_n|}(\bar{\Upsilon}^\times_{n-1}
    +\bar{\Upsilon}^\times_{n})\right|+|\phi_n^{\{\pm\}}||\daleth_n-\daleth_{n-1}|
    +2|\daleth_{n-1}|\phi_n^{\{\pm\}}-\phi_{n-1}^{\{\pm\}}| \label{4.84}
  \end{eqnarray}
  Inequalities $(\ref{4.44})-(\ref{4.46})-(\ref{4.48})-(\ref{4.50})$ 
give  bounds of the right hand side of $(\ref{4.84})$ except the term   $(\bar{\Upsilon}_{n}-\bar{\Upsilon}_{n-1})$.
 Using (\ref{4.12})-(\ref{4.13}) and (\ref{4.83}), one have
  \begin{equation}\label{4.85}
    (\bar{\Upsilon}_{n+1}-\bar{\Upsilon}_{n})(t_k,\theta_k) \leq
    C+C'\int_{t_0}^{t_k}(\bar{\Upsilon}_{n}-\bar{\Upsilon}_{n-1})dt
  \end{equation}
 Using $(\ref{4.56'})$ in $(\ref{23})-(\ref{24})$ gives for \ $~j,l\in\{1;2;3\}$,
  \begin{eqnarray}
   |\rho_{_{l,n}}-\rho_{_{l,n-1}}|,~|S_{lj,n}-S_{lj,n-1}|,~|P_{l,n}-P_{l,n-1}|&\lesssim&~~~~~\nonumber \\
16 \pi
    \|f_{n+1}-f_n\|_{(L^1(\mathbb{R}^3(dv)))}+\|f_0\|_\infty|\alpha_n-\alpha_{n-1}|\bar{u}_{n,1}ln(\bar{u}_{n,1})
    ~&~&~~~~~\label{4.85+}
  \end{eqnarray}
  Relation $(\ref{2+})$ and the mean value theorem give
  \begin{equation}\label{4.85++}
    |(f_{n+1}-f_n)(t)|\leq C  |\partial_{(\theta,v)} f_0|
   \int_{t_0}^t\digamma_{n-1}(s)ds
\end{equation}
 We obtain from $(\ref{39ARWn})-(\ref{41ARWn})$ that
\begin{eqnarray}
|J_{n+1}-J_n|(t) =
2t\left|\int_{t_0}^t(e^{\beta_n}S_{12,n}-e^{\beta_{n-1}}S_{12,n-1})ds\right|
\label{4.86}\\
|K_{n+1}-K_n|(t) \leq
2t^2\left|\int_{t_0}^t[e^{\beta_n}A_nS_{12,n}-e^{\beta_{n-1}}A_{n-1}S_{12,{n-1}}]ds\right|\nonumber\\
 +2t^3\left|\int_{t_0}^t[e^{\beta_n}e^{-2\mu_n}S_{13,n}-e^{\beta_{n-1}}e^{-2\mu_{n-1}}S_{13,{n-1}}]ds\right|
\label{4.87}
\end{eqnarray}
 In other ways we have
\begin{eqnarray}
\int_{t_0}^t[e^{4\mu_n}\Gamma_n^2-e^{4\mu_{n-1}}\Gamma_{n-1}^2](s,\theta)ds&\leq& \int_{t_0}^t\int_{\bar{\theta}}^\theta |e^{4\mu_{n}}\Gamma_{n}^2(2\beta'_n-4\mu'_n)\nonumber  \\
  -e^{4\mu_{n-1}}\Gamma_{n-1}^2(2\beta'_{n-1}-4\mu'_{n-1})|(s,\acute{\theta})d\acute{\theta}ds&+&\int_{t_0}^t[e^{4\mu_n}
  \Gamma_n^2-e^{4\mu_{n-1}}\Gamma_{n-1}^2](s,\bar{\theta})ds  \nonumber
\end{eqnarray}
and
\begin{eqnarray*}
 &~& |e^{\beta_n}(K_n-A_nJ_n)-e^{\beta_{n-1}}(K_{n-1}-A_{n-1}J_{n-1})|(t,\theta)
  \leq  |e^{\beta_n}(K_n-A_nJ_n) \\
  ~&~&-e^{\beta_{n-1}}(K_{n-1}-A_{n-1}J_{n-1})|(t,\bar{\theta})
  +\int_{\bar{\theta}}^\theta[|e^{\beta_{n}}-e^{\beta_{n-1}}||\beta'_n(K_n-A_nJ_n)|\\
  & + & e^{\beta_{n-1}}|\beta'_n(K_n A_n J_n)-\beta'_{n-1}(K_{n-1}-A_{n-1}J_{n-1})|](t,\acute{\theta})d\acute{\theta}
\end{eqnarray*}
 which implies
 \begin{equation*}
 \int_{t_0}^t|\dot{H}_n-\dot{H}_{n-1}|(s,\theta)ds \leq \int_{t_0}^t[e^{\tau_n}|\dot{H}_{n-1}(e^{-\tau_n}-e^{-\tau_{n-1}})|+\digamma_n](s,\theta)ds
\end{equation*}
and then
\begin{equation*}
|H_n-H_{n-1}|(t,\theta)\lesssim \int_{t_0}^t
\digamma_n(s, \theta)ds
\end{equation*}
Adding $(\ref{4.5})-(\ref{4.6})$ and use $(\ref{betat})$ gives
\begin{equation}\label{4.88}
|\dot{\beta}_n-\dot{\beta}_{n-1}|=\left|t(\bar{\Upsilon}_n-\bar{\Upsilon}_{n-1})+(\Gamma_n-\Gamma_{n-1})+
(\dot{H}_n-\dot{H}_{n-1})+8\pi(P^f_{1,n}-P^f_{1,n-1})\right|
\end{equation}
Adding
$(\ref{4.85})-(\ref{4.86})-(\ref{4.87})-(\ref{4.88})~and~(\ref{4.85++})$,
implies
\begin{equation}\label{4.89}
    \digamma_{n}(t)\leq
    C\int_{t_0}^t(\digamma_{n}(s)+\digamma_{n-1}(s))ds;~~~~n\geq 1
\end{equation}
Applying Gronwall's inequality  in $(\ref{4.89})$ show that
\begin{equation}\label{4.90}
\digamma_{n}(t)\leq
    C\int_{t_0}^t\digamma_{n-1}(s)ds
\end{equation}
 and by induction
 \begin{equation*}
 \digamma_{n}(t)\leq C^{n+1}\frac{(t-t_0)^n}{n!}~~~~\text{for n}
 \in \mathbb{N}, \ ~ and ~~~t\in [t_0,T^*]
 \end{equation*}
This proves that $\digamma_n\rightarrow 0~when~n\rightarrow
+\infty$ since the series
$\left\{C^{n+1}\frac{(t-t_0)^n}{n!}\right\}_{n\in \mathbb{N}^*}$ converges. \\
We conclude that there exist functions $(\alpha,~A_,~\mu,~\tau,~\phi,~f,~H,~G)$ solution of the complete system  such that
$(\alpha_n,~~A_n,~\mu_n,~\tau_n,~\phi_n,~f_n,~H_n,~G_n)$ converges to $(\alpha,~A_,~\mu,~\tau,~\phi,~f,~H,~G)$ in $L^\infty-$ norm.\\
Let us now prove uniqueness. Suppose
\begin{equation*}
\chi_k=(f_k;~\alpha_k;~\mu_k;~A_k;~\tau_k;~G_k;~H_k;~\phi_k)_{k=1;2}
\end{equation*}
are two regular solutions of the Cauchy problem
for the same initial data.
Then we deduce from (\ref{4.90}) that
\begin{equation*}
\daleth(t)\leq
    C\int_{t_0}^t\daleth(s)ds
\end{equation*}
where
\begin{eqnarray*}
 ~\daleth(t)&:=&\sup\{\|(f_{1}-f_2)(s)\|+\|(\phi_{1}-\phi_2)(s)\|+\|(\alpha_{1}-\alpha_2)(s)\|\nonumber \\
   ~ &+& \|(\mu_{1}-\mu_2)(s)\||+\|(A_{1}-A_2)(s)\|+\|(\tau_{1}-\tau_2)(s)\| \nonumber  \\
   ~ &+&
   \|(G_{1}-G_2)(s)\|+\|(H_{1}-H_2)(s)\|;t_0\leq s
   \leq T^*\}  
 \end{eqnarray*} 
 This implies that $\daleth(t)=0~for~t\in[t_0;T^*[$.
We have proved the following result:
\begin{theorem}
{\bf(local existence)}
Let $\overset{\circ}{f} \in C^{1}(S^1\times \mathbb{R}^{3})$
with $\overset{\circ}{f}\geq 0$, \\ $\overset{\circ}{f}(\theta+2\pi,v) = \overset{\circ}{f}(\theta, v)$
for $(\theta,v) \in S^1\times\mathbb{R}^{2}$ and \\
$F_{0} := \sup \{ |v| | (\theta,v) \in {\rm supp} \overset{\circ}{f}\} <\infty$. \
Let given regular functions $\bar{\alpha}, \bar{\tau}, \bar{\mu}, \bar{A}, \bar{H}, \bar{G}, \psi \in C^{1}(\mathbb{R})$ and $\overset{\circ}{\alpha}, \overset{\circ}{\tau}, \overset{\circ}{\mu}, \overset{\circ}{A}, \overset{\circ}{H}, \overset{\circ}{G}, \overset{\circ}{\phi} \in C^{2}(S^1)$.
Then there exists a unique right maximal regular solution $(\alpha,
\tau, \mu, A, H, G, f, \phi)$ of the complete EVSFS   with
$(\alpha, \tau, \mu, A, H,\\ G, f, \phi)(t_0) = (\overset{\circ}{\alpha},
\overset{\circ}{\tau}, \overset{\circ}{\mu}, \overset{\circ}{A}, \overset{\circ}{H},
\overset{\circ}{G}, \overset{\circ}{f},
\overset{\circ}{\phi})$ and $(\dot{\alpha}, \dot{\tau}, \dot{\mu}, \dot{A}, \dot{H},
\dot{G},\dot{\phi})(t_0) = ( \bar{\tau}, \bar{\mu}, \bar{A}, \bar{H}, \bar{G}, \psi)$ on a time
interval $[t_0,T[$ with $T \ge t_0$.
\end{theorem}
\begin{remark}
The result  obtained here proves  particularly the local in time existence of solution of Einstein-Vlasov system
in the case of Gowdy symmetry with scalar field or not in future direction. It proves  also local in time existence of
solution in the $T^2$ symmetry case without scalar field.
It remains following $[L]{}$, establishing global existence result and looking for stability  and asymptotic behaviour of that solution (cf $[R]{}$, $[D]{}$).
\end{remark}
This research did not received any specific grant from funding agencies in the public, commercial, or not-for-profit sectors.
\section*{References}
$[A]{}$ H. Andreasson,  Global foliations of matter spacetimes with Gowdy symmetry.\ \ Comm.Math.Phys. $206, \ pp. 337-365, \ (2003)$.\\
$[A1]{}$ H. Andreasson, The Einstein-Vlasov system. Kinetic Theory, Living Rev. Relativity, $5, ~7,~ (2014)$.\\
$[ARW]{}$ H. Andreasson, A. D. Rendall, and M. Weaver, Existence of CMC and constant areal time foliations in $T^2$-symmetric spacetimes with Vlasov matter. Comm. Partial Differential Equations,
$29, N^o.1-2, \ pp. 237-262,~(2004)$. \\
$[BCIM]{}$ B.K. Berger, P. Chrusciel, J. Isenberg and V. Moncrief, Global foliations
of vacuum spacetimes with $T^2$ isometry. Ann. Phys. $260, pp.  117-148, \ (1997)$.\\
$[C]{}$ Y. Choquet-Bruhat, Probl\`eme de Cauchy pour le syst\`eme int\'egro-diff\'erentiel
d'Einstein-Liouville, Ann. Inst. Fourier, tome 21, $N^03$ \ pp. 181-201 (1971).\\
$[D]{}$  M. Dafermos and A.D. Rendall, Strong cosmic censorship for surface-symmetric cosmological spacetimes with collisionless matter, Comm. Pure and Applied Math. Vol. 69 (5) : https://doi.org/10.1002/cpa.21603 \ (2016).\\ 
$[L]{}$ A.T. Lassiye, \'Equations d'Einstein-Vlasov en sym\'etrie  $T^2$ avec champ
scalaire non lin\'eaire. Th\`{e}se de Doctorat/Ph.D en Math\'ematiques, Universit\'e de Yaounde 1 $(2020)$.\\
$[JLS]{}$ J. Luk, R. M. Strain, Strichartz estimates and moment bounds for the relativistic Vlasov-Maxwell
system II. Continuation criteria in the $3$D case. Archive for Rational Mechanic and Analysis 219, \ pp. 445-552 (2016). 
$[R]{}$ H. Ringstr$\ddot{o}$m, Future stability of the Einstein non-linear scalar field system. Invent. Math. $173, 123-208, (2008)$.\\
$[S]{}$ J. Smulevici, On the area of the symmetry orbits of cosmological spacetimes with teroidal or hyperbolic symmetry.
Anal. PDE $4(2), \ 91-245 \ (2011)$.\\ 
$[T1]{}$ D. Tegankong, Cosmological solutions of the
Einstein-Vlasov-scalar field system.\ \ Journ. Hyperb. Diff .Eq. Vol $1,N^o.4, \ pp.691-624,~(2005)$. \\
$[T2]{}$ D. Tegankong, The Einstein-Vlasov scalar field system with Gowdy symmetry in the expanding direction.
Classical Quantum Gravity $31 ~,~ no.~ 15,~ 155008,~ 18.\\ ~ MR~3233262~(2014)$.\\
$[W]{}$ M. Weaver,  On the area of the symmetry orbits in $T^2$-symmetric spacetimes with Vlasov matter. Classical Quantum Gravity, $21 , ~pp.1079-1097,~(2004)$.

\end{document}